\documentclass[twocolumn,english,amsmath,amssymb,aps,prl,superscriptaddress]{revtex4-2}
\usepackage[T1]{fontenc}
\usepackage[latin9]{inputenc}
\usepackage{babel}
\usepackage{float}
\usepackage{amsmath}
\usepackage{graphicx}
\usepackage{colortbl}
\usepackage{relsize}

\usepackage[unicode=true,
 bookmarks=false,
 breaklinks=true,pdfborder={0 0 1},backref=false,colorlinks=true]
 {hyperref}
\hypersetup{pdfcreator={},pdfproducer={LaTeX with hyperref},linkcolor=blue,anchorcolor=blue,citecolor=blue,filecolor=red,menucolor=red,urlcolor=blue,pdfstartview=FitV,pdfhighlight=/I,hypertexnames=true}

\makeatletter
\usepackage{natbib}
\setcitestyle{numbers}

\makeatother

\begin{document}
\title{Orbital altermagnetism on the kagome lattice and possible application to $A$V$_3$Sb$_5$}
\author{Anzumaan R. Chakraborty}
\affiliation{Department of Physics, The Grainger College of Engineering, University of Illinois Urbana-Champaign, Urbana, IL 61801, USA}
\affiliation{Anthony J. Leggett Institute for Condensed Matter Theory, The Grainger College of Engineering, University of Illinois Urbana-Champaign, Urbana, 61801, IL, USA}

\author{Fan Yang}
\affiliation{Department of Chemical Engineering and Materials Science,
University of Minnesota, Minneapolis, Minnesota 55455, USA}

\author{Turan Birol}
\affiliation{Department of Chemical Engineering and Materials Science,
University of Minnesota, Minneapolis, Minnesota 55455, USA}

\author{Rafael M. Fernandes}
\affiliation{Department of Physics, The Grainger College of Engineering, University of Illinois Urbana-Champaign, Urbana, IL 61801, USA}
\affiliation{Anthony J. Leggett Institute for Condensed Matter Theory, The Grainger College of Engineering, University of Illinois Urbana-Champaign, Urbana, 61801, IL, USA}

\date{\today}
\begin{abstract}
Altermagnets, which encompass a broad landscape of materials, are compensated collinear magnetic phases in which the antiparallel magnetic moments are related by a crystalline rotation. Here, we argue that collinear altermagnetic-like states can also be realized in lattices with an odd number of sublattices, provided that the electronic interactions promote non-uniform magnetic moments. We demonstrate this idea for a kagome metal whose band filling places the Fermi level close to the van Hove singularity. Combining phenomenological and microscopic modeling, we show that the intertwined charge density-wave and loop-current instabilities of this model lead to a wide parameter range in which orbital ferromagnetic, antiferromagnetic, and altermagnetic phases emerge inside the charge-ordered state. In the presence of spin-orbit coupling, their electronic structures display the usual spin-split fingerprints associated with the three types of collinear magnetic order. We discuss the possible realization of orbital altermagnetic phases in the $A$V$_3$Sb$_5$ family of kagome metals.
\end{abstract}
\maketitle

\section*{Introduction}
The recent classification of collinear magnetic states via spin groups has revealed a broad class of compensated magnets called altermagnets (AM) \cite{Smejkal2020_AM,Smejkal2022_AM,Smejkal2022_2_AM}, whose properties are intermediate between those of ferromagnets (FM) and N\'eel antiferromagnets (AFM). Invariant under a combination of time reversal and crystal rotation, AM exhibit zero net magnetization alongside spin-split band dispersions of nodal $d$-, $g$-, or $i$-wave character \cite{Jungwirth2025symmetry}. Their discovery sparked intense interest due to their relevance to spintronics \cite{dal2024antiferromagnetic,jungwirth2025altermagnetic}, multipolar magnetism \cite{Hayami2018_multipoles,Voleti2020,Bhowal2024,Mcclarty2024}, electronic correlations \cite{Jungwirth_Fernandes2025,Leeb2024,Yu_Agterberg2024,Valenti2024,Kaushal2024,Sobral2024,Giuli2025,Thomale2025}, and topology \cite{Cano2024,Fernandes2024_AM,Knolle2024,Mazin2023_FeSe,Attias2024,Schnyder2024,Agterberg2024,Antonenko2025}.

The formal definition provides a straightforward blueprint for constructing AM states: on non-Bravais lattices with an even number of sublattices not related by inversion, place magnetic moments pointing up on half the sublattices and down on the other half \cite{Schiff2024,Smolyanyuk2024}. This applies to materials like CrSb \cite{Reimers2023,Ding2024,Yang2025,Li2024} and MnTe \cite{Amin2024,Krempasky2024,Lee2023,Osumi2023}, where sublattices are related by a screw rotation, and to the AM candidate KMnF$_3$ \cite{naka2025altermagnetic}, with four sublattices related by a glide plane. Crucially, this suggests AM cannot form in crystals with an odd number of magnetic sublattices, as a collinear compensated state would be impossible. A potential workaround involves non-collinear configurations, like the $120^\circ$ phase of Mn$_3$Sn \cite{sticht1989non}. However, this phase exhibits non-collinear spin-textured band dispersions rather than bands spin-polarized along the same axis across the Brillouin zone \cite{Liu2024,Fang2024,Song2024}.

\begin{figure}
\includegraphics[width=.8\columnwidth]{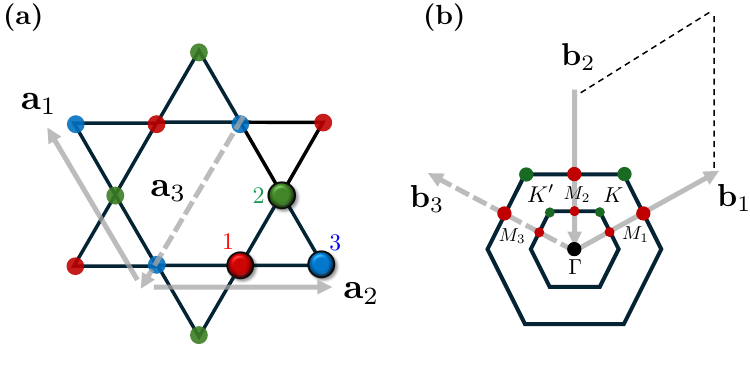}
\caption{\textbf{(a)} The kagome lattice vectors $\textbf{a}_{1,2,3}$ and the sublattices $1$ (red), $2$ (green), and $3$ (blue). \textbf{(b)} Brillouin zone with reciprocal vectors $\textbf{b}_{1,2,3}$ and the high-symmetry points $\Gamma$ (black), K (green), and M (red). The smaller hexagon is the Brillouin zone in the CDW phase; note that the M points of the larger BZ are folded onto the $\Gamma$ point.} \label{fig:lattice}
\end{figure}

In this paper, we show that nodal collinear AM-like states exist in crystals with an odd number of sublattices when non-uniform collinear magnetic configurations are allowed -- that is, when the magnetic moment amplitudes vary across sublattices. Consider, for example, the kagome lattice shown in Fig. \ref{fig:lattice}(a). Placing an up moment on sublattice 1, a down moment on sublattice 3, and zero moment on sublattice 2 yields a $d$-wave AM state. While less common than uniform configurations, non-uniform magnetic states are well known in correlated systems \cite{Lorenzana2008,Nandkishore2012,Fernandes2016}. Experimentally, a collinear magnetic state in which half the magnetic sites have zero magnetization was observed in hole-doped iron pnictides \cite{Allred2016}. Other examples of non-uniform magnetic states where some sites show no magnetic moments are the AM candidate Mn$_5$Si$_3$ \cite{reichlova2024observation} and the kagome AFM Fe$_4$Si$_2$Si$_7$O$_{16}$ \cite{Ling2017Striped}. More broadly, in multi-orbital systems, the competition between finite and zero spin states may also stabilize non-uniform states \cite{Kruger2009,Khaliullin2013}. 

Beyond spin systems, loop currents typically induce magnetic moments with varying amplitudes within the unit cell \cite{Fernandes2025}. Here, we show that itinerant electrons on a kagome lattice near the van Hove singularity (vHs) realize a non-uniform collinear orbital magnetic state with the symmetries of a $d$-wave AM. In this regime, electron-electron interactions favor competing charge-density wave (CDW) and loop-current (LC) orders at the $M=(1/2,\,1/2)$ wave-vector (Fig. \ref{fig:lattice}(b)) \cite{Park2021,Lin2021,Denner2021,christensen2021theory,JPHu2021,christensen2022toroidal,Tazai2022,Ferrari2022,Wu2023,Scammell2023,Fischer2023,SBLee2023,Ziqiang2023,HYKee2024,Ingham2025,Zhan2025}. Because of the anharmonic coupling between these order parameters \cite{christensen2022toroidal}, there is a wide regime in which CDW appears first, followed by LC order at lower temperature. In this case, since translational symmetry is already broken by CDW order, the LC phase emerges as a zero wave-vector orbital magnetic state. We stress that ``orbital'' here refers to the orbital motion of the electrons in the interatomic loop currents, and not to the atomic-orbital quantum numbers of the electrons. Such magnetic phases have also been known as flux phases or, more broadly, loop-current phases. \cite{hsu1991,varma1997,chakravarty2001,sun2008,Fernandes2025}.

We identify three leading LC instabilities from within the CDW state, corresponding to orbital AFM, FM, and $d$-wave AM. We compute their electronic spectra in the presence of spin-orbit coupling (SOC), which communicates the symmetries of the LC-induced orbital magnetic moments to spin degrees of freedom. Indeed, we find that the resulting band structures display no spin splitting (orbital AFM), net $s$-wave spin splitting (orbital FM), and $d$-wave spin splitting (orbital AM). A kagome material that has been proposed to realize LC order is the family of  $A$V$_3$Sb$_5$ kagome metals \cite{Ortiz2019,Ortiz2020,Ortiz2021,Kang2022,Jiang2021,mielke2022time,xing2024optical}. A candidate LC order, dubbed congruent CDW flux phase, was recently proposed based on scanning tunneling microscopy (STM) results. Interestingly, we find that our orbital $d$-wave altermagnetic state is the 2D version of this phase. As a result, we propose spin-resolved angle-resolved photo-emission spectroscopy (ARPES) as an ideal probe to unambiguously assess the realization of the congruent CDW flux phase in $A$V$_3$Sb$_5$. More broadly, our results extend AM to systems with an odd number of magnetic sublattices.

\section*{Intertwined CDW and LC orders}
In two dimensions, saddle points in the electronic dispersion yield a logarithmic divergence in the density of states, dubbed van Hove singularity (vHs). In the kagome lattice, the vHs sit at the three M-points $M_{1,2,3}$ related by three-fold rotations and with wave-vectors $\textbf{Q}_i\equiv \textbf{b}_i/2$ , where $\textbf{b}_{1,3}=\frac{2\pi}{\sqrt{3}}(\pm\sqrt{3},1)$ and $\textbf{b}_2=-\frac{4\pi}{\sqrt{3}}(0,1)$. The corresponding lattice vectors are given by $ \textbf{a}_{1,3}=\frac{1}{2}(-1,\pm\sqrt{3})$ and $\textbf{a}_2=(1,0)$, as shown in Fig. \ref{fig:lattice}. Notice that, in this convention, $\textbf{b}_i\cdot\textbf{a}_i=0$. 

When the chemical potential is near the vHs, as is the case for $A$V$_3$Sb$_5$ \cite{Kang2022,Wilson2024}, electron-electron interactions favor particle-hole instabilities with wave-vector $\textbf{Q}_i$ \cite{Kiesel2012,Nandkishore2012}.  In the charge channel, these instabilities either preserve time-reversal symmetry (TRS), corresponding to a charge density-wave (CDW), or break it, corresponding to a loop current (LC) state. The former creates a pattern of dimerized nearest-neighbor bonds, whereas the latter is manifested as microscopic currents along the bonds that give rise to orbital moments at the centers of the plaquettes. 

Various regimes of electron-electron interactions have been reported that favor competing CDW and LC instabilities on the kagome lattice \cite{Park2021,Lin2021,Denner2021,christensen2021theory,JPHu2021,christensen2022toroidal,Tazai2022,Ferrari2022,Wu2023,Scammell2023,Fischer2023,SBLee2023,Ziqiang2023,HYKee2024,Ingham2025,Zhan2025}. To encompass this broad landscape of possible microscopic mechanisms, we employ a phenomenological approach and construct the CDW and LC order parameters $W_i$ and $\Phi_i$, respectively, with wave-vector $\textbf{Q}_i\equiv \textbf{b}_i/2$  \cite{christensen2022toroidal,Fischer2023}.  In terms of the fermionic operator $d_{j\textbf{r},\alpha}$ that annihilates an electron at Bravais lattice vector $\textbf{r}$, sublattice $j$, and spin projection $\alpha \in {\{\uparrow,\downarrow\}}$, they are given by:
\begin{align}
   W_i  & = \sum\langle e^{\mathrm{i}\textbf{Q}_i\cdot \textbf{r}}d_{j\textbf{r},\alpha}^\dagger(d_{l\textbf{r},\alpha}^{\phantom{}}-d_{l\textbf{r}+\textbf{a}_i,\alpha})+\text{H.c.}\rangle \\
    \Phi_i & = \mathrm{i} \sum\langle e^{\mathrm{i}\textbf{Q}_i\cdot \textbf{r}}d_{j\textbf{r},\alpha}^\dagger(d_{l\textbf{r},\alpha}^{\phantom{}}-d_{l\textbf{r}+\textbf{a}_i,\alpha})-\text{H.c.}\rangle \label{eq:OP}
\end{align}
Here,  $(i,j,l)$ is a cyclic permutation of $(1,2,3)$ defined such that the sublattices $j$ and $l$ in a single unit cell are connected by $\textbf{a}_i/2$, see Fig. \ref{fig:lattice}(a). In terms of the irreducible representations (irreps) of the space group $P6/mmm$ that describes the kagome lattice, the three-component CDW order parameter $\textbf{W}=(W_1,W_2,W_3)$ transforms as the irrep $M_1^+$, whereas the LC order parameter  $\boldsymbol{\Phi}=(\Phi_1,\Phi_2,\Phi_3)$ transforms as the irrep  $mM_2^+$  \cite{christensen2022toroidal}. Importantly, these irrep assignments do not depend on the symmetry of the orbital involved.

The CDW and LC instabilities have different bare critical temperatures $T_W^0$ and $T_\Phi^0$, respectively. However, the Landau free energy $\mathcal{F}(\textbf{W},\boldsymbol{\Phi})$ contains anharmonic terms that inevitably intertwine these two orders. 

While the full expression for $\mathcal{F}(\textbf{W},\boldsymbol{\Phi})$ was derived in Ref. \cite{christensen2022toroidal} and is repeated in the Supplemental Material (SM) \cite{SM} (see also references [6-8] therein), we illustrate this intertwining by analyzing the linear-quadratic term
\begin{equation}
     \gamma(W_1\Phi_2 \Phi_3+W_2\Phi_3 \Phi_1+W_3\Phi_1\Phi_2)
     \label{linearquadratic}
 \end{equation}
 
Because of this term, the condensation of triple-Q LC order necessarily triggers CDW order. Thus, even if $T_W^0 < T_\Phi^0$ at the bare (quadratic) level, the CDW transition temperature is renormalized to $T_W = T_\Phi^0$ . Conversely, in the case $T_W^0 > T_\Phi^0$, the condensation of $\textbf{W}$ enhances the LC transition temperature to $ T_\Phi - T_\Phi^0 \propto |\gamma| W$. In this work, we are interested on the latter case: because the onset of the triple-Q CDW with wave-vectors $\textbf{Q}_i\equiv \textbf{b}_i/2$ ($i=1,2,3$) quadruples the unit cell, the subsequent LC instability corresponds to a uniform magnetic state, since the LC wave-vector is folded onto $\textbf{Q}=0$. Because the magnetic moments arise from microscopic loop currents rather than spins, they often have different amplitudes on different plaquettes, resulting in non-uniform magnetic configurations.
\begin{figure*}
\begin{flushleft}
\includegraphics[width=2.0\columnwidth]{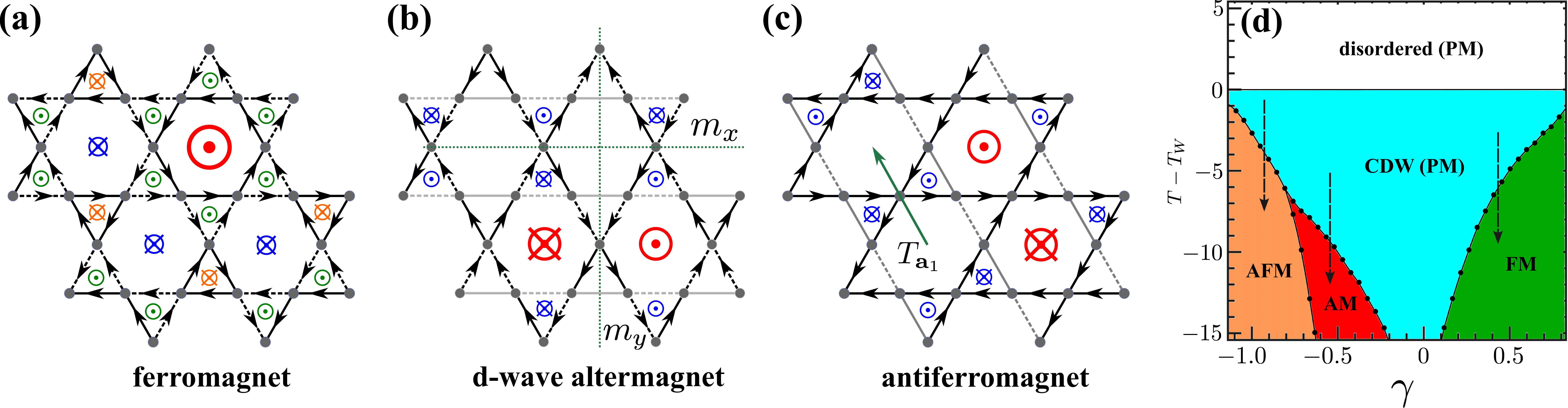}\caption{Bond dimerization and loop current patterns from the CDW-LC configurations associated with the ferromagnetic (FM) \textbf{(a)}, $d$-wave altermagnetic (AM) \textbf{(b)}, and antiferromagnetic (AFM) states \textbf{(c)}.The direction of the magnetic moment of a plaquette is determined by its net current circulation. Different colors indicate different magnitudes of the moments. \textbf{(d)}: Mean field phase diagram obtained from minimizing $\mathcal{F}(\textbf{W},\boldsymbol{\Phi})$  as a function of $T$ and $\gamma$. The phase boundaries are interpolated along discrete points (block dots) and the parameters used are listed in the SM. \label{fig:LC_states}}
\end{flushleft}
\end{figure*}

\textit{Phenomenological model.} --- When the CDW is the leading instability ($T_W^0 > T_\Phi^0$), the order parameter configuration that minimizes the free-energy is the triple-Q  $\textbf{W}=\pm W(1,1,1)$, known in the kagome lattice as star-of-David (plus sign) or tri-hexagonal (minus sign), both of which quadruple the unit cell \cite{christensen2021theory}.

As a result, a LC instability inside this triple-Q CDW phase  corresponds to a uniform (i.e.,  $\textbf{Q}=0$) magnetic order. To see this, we form bilinears between $\textbf{W}$ and $\boldsymbol{\Phi}$ exploiting the irrep decomposition
\begin{equation}
    M_1^+ \otimes mM_2^+ =m\Gamma_2^+ \oplus m\Gamma_5^+\oplus mM_1^+\oplus mM_2^+\,.\label{eq:decomposition}
\end{equation}
We thus introduce two uniform magnetic order parameters: the single-component $\mathcal{M}$, which transforms as an out-of-plane orbital ferromagnetic (FM) moment (irrep $m\Gamma_2^+$ of the space group or, equivalently, $A_{2g}^-$ irrep of the point group  $D_{6h}$) and the two-component  $\boldsymbol{\mathcal{N}}=(\mathcal{N}_1,\mathcal{N}_2)$. The latter transforms as the irrep $ m\Gamma_5^+$ ($E_{2g}^-$ irrep of the point group  $D_{6h}$), which is odd under time-reversal symmetry and breaks threefold rotational symmetry, and therefore corresponds to a $d$-wave orbital altermagnetic order parameter \cite{Fernandes2024_AM}. In the SM, we contextualize these non-uniform magnetic states using the classifications of magnetic states in terms of spin and magnetic groups. Inside the CDW phase, where $W\neq0$ can be approximated as a constant, the three components of $\boldsymbol{\Phi}$ are related to $ \mathcal{M}$ and $\boldsymbol{\mathcal{N}}$, to leading order, via: 

\begin{equation}
\left(\begin{array}{c}
\mathcal{M}\\
\mathcal{N}_{1}\\
\mathcal{N}_{2}
\end{array}\right)=W\left(\begin{array}{ccc}
1 & 1 & 1\\
\sqrt{3} & 0 & -\sqrt{3}\\
1 & -2 & 1
\end{array}\right)\left(\begin{array}{c}
\Phi_{1}\\
\Phi_{2}\\
\Phi_{3}
\end{array}\right) \, .\label{eq:transformation}
\end{equation}
We now use this linear transformation to recast $\mathcal{F}(\textbf{W},\boldsymbol{\Phi})$ for a constant $\textbf{W}=-W(1,1,1)$ in terms of an effective free energy for the uniform order parameters:
\begin{equation}
\label{compositeLandau}
    \begin{split}
    \mathcal{F}_{\mathrm{eff}}(\mathcal{M},\boldsymbol{\mathcal{N}}) &= \frac{a_1}{2}\mathcal{M}^2+\frac{u_1}{4}\mathcal{M}^4 \\
    & +\frac{a_2}{2}\mathcal{N}^2+\frac{u_2}{4}\mathcal{N}^4+\frac{\zeta}{6}\mathcal{N}^6\cos6\theta \\
    & +\frac{u}{4} \mathcal{M}^2\mathcal{N}^2+\frac{\lambda}{4}\mathcal{M}\mathcal{N}^3\sin 3\theta \, ,
\end{split}
\end{equation}
where  $\boldsymbol{\mathcal{N}}\equiv\mathcal{N}(\cos\theta,\sin\theta)$. The Landau coefficients of $\mathcal{F}_{\mathrm{eff}}$ are given in terms of the coefficients of $\mathcal{F}$ in the SM. Setting the quadratic coefficients to zero yields the transition temperatures $T_\mathcal{M}=T_{\Phi,0}+2W\gamma/3$ and $T_\mathcal{N}=T_{\Phi,0}-W\gamma/3$ to leading order in $W$. Hence, the state favored inside the CDW phase depends on the sign of $\gamma$ in Eq. (\ref{linearquadratic}), with  $\gamma>0$ ($\gamma<0$) favoring FM (AM) order. 

The orbital FM and AM states are illustrated in Figs. \ref{fig:LC_states}(a) and (b), respectively. By inverting Eq. (\ref{eq:transformation}), the $\mathcal{M}\neq0$, $\mathcal{N}=0$ FM state corresponds to the order parameter configuration $\textbf{W}=-W(1,1,1)$ and $\boldsymbol{\Phi}=\Phi(1,1,1)$. As shown in Fig. \ref{fig:LC_states}(a), the LC pattern breaks the vertical mirrors, and thus each plaquette develops a nonzero out-of-plane orbital magnetic moment at its center.  The different colored circles represent moments with different magnitudes, reflecting their distinct local crystalline environments caused by the CDW bond distortions. Because of this, as discussed in Ref. \cite{christensen2022toroidal}, these orbital moments do not cancel across the quadrupled unit cell of the CDW phase, leading to a net ferromagnetic moment. 

As for the $d$-wave AM state with $\boldsymbol{\mathcal{N}}=\mathcal{N}(\cos\theta,\sin\theta)\neq0$, there are two distinct possibilities depending on the values of $\theta$  that minimize the sixth-order term in Eq. (\ref{compositeLandau}) with Landau coefficient $\zeta$. There are six possible values $\theta_n= n \pi/6$ with even $n=2m$ for $\zeta<0$  and odd $n=2m+1$ for $\zeta>0$, where $m=0,1,...,5$. The first case corresponds to a ``pure'' $d$-wave AM phase \cite{Fernandes2024_AM}, which triggers secondary nematic and $i$-wave AM order parameters (see SM). From Eq. (\ref{eq:transformation}), the LC configuration $\boldsymbol{\Phi}$ corresponding to this AM state only has two non-zero components with opposite signs and same magnitude. 

\begin{figure*}
\includegraphics[width=2.0\columnwidth]{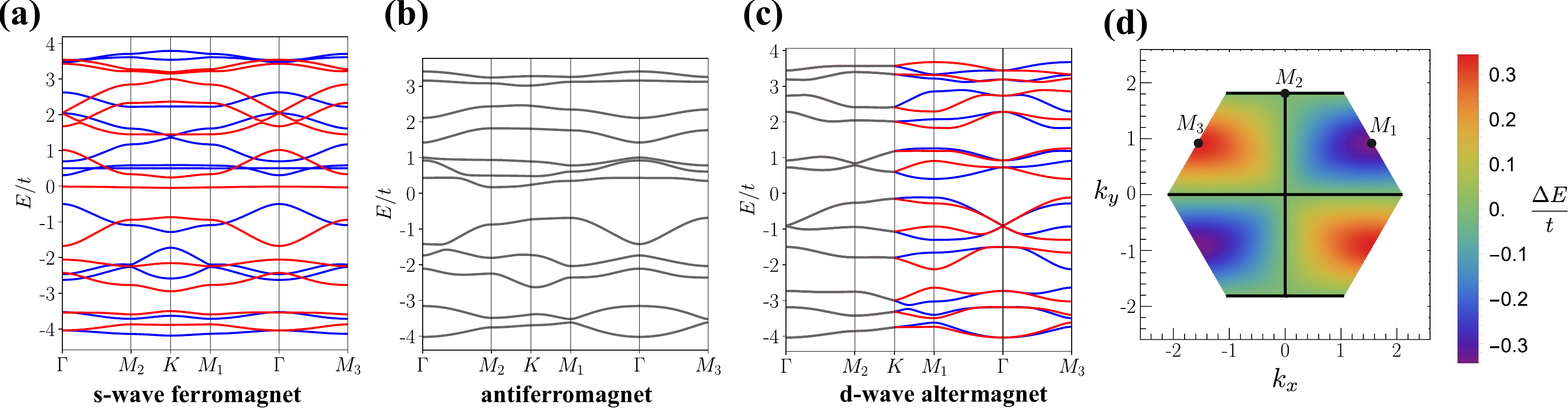}\caption{Electronic spectrum of the FM \textbf{(a)}, AFM \textbf{(b)}, and AM \textbf{(c)} phases shown in Fig. \ref{fig:LC_states} obtained by diagonalizing $\mathcal{H}_\text{tot}$. Red and blue denote spin-up and spin-down bands. The $\textbf{k}$-space points refer to the smaller BZ of the $2\times2$ unit cell (Fig. \ref{fig:lattice}(b)). The parameters used are listed in the SM. \textbf{(d)} Spin-splitting $\Delta E(\textbf{k})\equiv E_\uparrow - E_\downarrow$ of the highest-energy band of the AM phase plotted along the entire BZ.}   \label{fig:bands}
\end{figure*}

Fig. \ref{fig:LC_states}(b) illustrates the LC configuration $\boldsymbol{\Phi}=\Phi(1,0,-1)$ inside the $\textbf{W}=-W(1,1+\delta _w,1)$ CDW state. Note that the non-zero $\delta_w$ is induced inside the AM phase due to the anharmonic coupling in Eq. (\ref{linearquadratic}). As shown in the figure, out-of-plane orbital magnetic moments with two different magnitudes emerge in some of the plaquettes (red and blue circles). In contrast to the FM case, these moments cancel out, implying a compensated magnetic state. Importantly, each pair of orbital magnetic moments with opposite directions is related by one of the vertical mirrors $m_x$ and $m_y$ (dashed green lines) within the quadrupled unit cell. This further demonstrates that this is an altermagnetic phase.

The other possible $d$-wave AM phase is characterized by $\theta_n= n \pi/6$ with odd $n$.  The main difference is that it also induces a net out-of-plane orbital magnetic moment, as can be seen from the last term in the Landau free energy of Eq. (\ref{compositeLandau}). As a result, this is not a ``pure'' AM phase, and will not be discussed in the remainder of this paper.

While the analysis above is restricted to LC states that coexist with the triple-Q CDW order, a group-theory analysis reveals four additional symmetry-distinct mixed CDW-LC states (see SM). One of them, which is particularly favored by the anharmonic coupling  in Eq. (\ref{linearquadratic}), is described by $\textbf{W}= W(1,0,0)$ and $\boldsymbol{\Phi}=\Phi(0,1,1)$ \cite{christensen2022toroidal}. In contrast to the AM and FM phases, this state does not have any finite zero-momentum bilinear related to the decomposition (\ref{eq:decomposition}).  The corresponding configuration of loop-currents and bond-dimerization is shown in Fig. \ref{fig:LC_states}(c). As in Fig. \ref{fig:LC_states}(b), threefold rotational symmetry is broken and out-of-plane orbital magnetic moments with different magnitudes emerge. In contrast to the AM case, however, pairs of opposite moments are related by a translation by $\textbf{a}_1$ (green arrow). Consequently, this is an orbital antiferromagnetic state (AFM), which can only be achieved from the triple-Q CDW phase via a first-order transition.

To establish which of the three magnetic phases (FM, AM, and AFM) is realized, one must numerically minimize the free energy
$\mathcal{F}(\textbf{W},\boldsymbol{\Phi})$, which has a large parameter space. Fortunately, a free energy with the same form was systematically studied numerically in Ref. \cite{christensen2021theory}, where the role of $\boldsymbol{\Phi}$ was played by another CDW order parameter with out-of-plane wave-vector component. The phase diagrams obtained in that work in the regime $T_W^0 > T_\Phi^0$ revealed the overall predominance of three phases with $\boldsymbol{\Phi}\neq0$, which, in our case where $\boldsymbol{\Phi}$ is LC order, correspond precisely to the FM, AM, and AFM phases. Since our symmetry analysis suggested that the anharmonic coupling $\gamma$ from Eq. (\ref{linearquadratic}) can tune between these three phases, we obtained the $\gamma-T$ phase diagram from $\mathcal{F}(\textbf{W},\boldsymbol{\Phi})$. For concreteness, we used the same representative Landau coefficients values studied in Ref. \cite{christensen2021theory} and compared the energies of the three magnetic states and of the triple-Q tri-hexagonal CDW state. The result, shown in Fig. \ref{fig:LC_states}(d), reveals that, as expected, $\gamma>0$ favors FM order whereas $\gamma<0$ favors AM order inside the CDW phase. Upon further increasing the magnitude of $\gamma$ along the negative axis, the AFM state is stabilized. Thus, not only do the AFM, FM, and AM orders emerge inside the CDW state in our model, but also the parameter $\gamma$ tunes between different phases. We stress that the hierarchy of phases as a function of $\gamma$ in Fig. \ref{fig:LC_states}(d) is stable for a broad range of Landau parameters, as discussed in Ref. \cite{christensen2021theory} and extensively in the SM.

\section*{Microscopic model}
The key signature of altermagnetism is the nodal even-parity spin-splitting of the bands. Because our system has orbital magnetic moments, we must include spin-orbit coupling (SOC) to see spin-splitting. We thus construct a microscopic Hamiltonian to obtain the electronic spectrum in the FM, AM, and AFM phases of  Fig. \ref{fig:LC_states}. The non-interacting part $ \mathcal{H}_0$ consists of electrons with nearest-neighbor hopping parameter $t$ on the kagome lattice:
\begin{equation}
    \mathcal{H}_0=\sum_{i,\textbf{r},\alpha}\big[(-t+\mathrm{i}\lambda_\text{soc}\sigma_z^{\alpha\alpha})d_{j\textbf{r},\alpha}^\dagger(d_{l\textbf{r},\alpha}^{\phantom{}}+d_{l\textbf{r}+\textbf{a}_i,\alpha}^{\phantom{}})+\text{H.c.}\big],
\end{equation}
with fermionic operators $d_{j\textbf{r},\alpha}$ defined as in Eq. (\ref{eq:OP}). Here, $\lambda_\text{soc}$ is the Kane-Mele-like SOC term arising from spin-dependent hopping between nearest neighbors \cite{kane2005}. The CDW and LC order parameters $\textbf{W}$ and $\boldsymbol{\Phi}$ appear in the Hamiltonian via $\mathcal{H}_\text{CDW}$ and $\mathcal{H}_\text{LC}$, through a mean-field coupling to the fermionic bilinears in Eq. (\ref{eq:OP}). They thus correspond, respectively, to modulations in the amplitude and in the phase of the nearest-neighbor hopping parameter. We also include the symmetry-allowed coupling between $\textbf{W}$ and the onsite energies of the three sublattices
\begin{equation}
    \mathcal{H}_\text{CDW}'= \eta \sum_{i,\textbf{r},\alpha}\left( W_ie^{\mathrm{i}\textbf{Q}_i\cdot\textbf{r}}d_{i\textbf{r},\alpha}^\dagger d_{i\textbf{r},\alpha}^{\phantom{}}+\text{H.c.}\right),
\end{equation}
where $\eta$ is a dimensionless constant that relates the bond distortion amplitude to the onsite energy shift amplitude caused by the CDW order. We diagonalize the full Hamiltonian $\mathcal{H}_\text{tot}=\mathcal{H}_0+\mathcal{H}_\text{LC}+\mathcal{H}_\text{CDW}+\mathcal{H}_\text{CDW}'$ in the  $\textbf{W}$ and $\boldsymbol{\Phi}$  configurations corresponding to the FM, AFM, and AM phases and show the spin-resolved 12-band electronic dispersion in Fig. \ref{fig:bands}. Note that the $\textbf{k}$-space path refers to the Brillouin zone of the quadrupled unit cell in the CDW phase (smaller hexagon of Fig. \ref{fig:lattice}(b)). 

As shown in Fig. \ref{fig:bands}(a), the spin-up and spin-down bands (red and blue) are split along all directions in the FM phase. Since the model, even in the ordered phases, preserves the in-plane mirror symmetry $m_z$ at all $k$-points, the SOC does not introduce any complications and the spin along $z$ is a good quantum number for all Bloch states. Interestingly, the splitting is not uniform, which we attribute to the non-uniform orbital magnetic moments in the $2\times2$ unit cell, see Fig. \ref{fig:LC_states}(a). In the AFM phase, shown in Fig. \ref{fig:bands}(b), all bands are Kramers degenerate, which is a consequence of the anti-unitary operator $t\mathcal{T}$ (time-reversal followed by translation) commuting with $\mathcal{H}$. In contrast, the spectrum of the AM phase displayed in Fig. \ref{fig:bands}(c) exhibits spin splittings with opposite signs along the orthogonal $\Gamma$-$M_1$ and  $\Gamma$-$M_3$ directions, and no spin splitting along the  $\Gamma$-$M_2$ direction, as expected for a $d$-wave altermagnet. Indeed, the full momentum-dependence of the spin splitting, shown in Fig. \ref{fig:bands}(d), is consistent with the symmetry properties of the AM state shown in Fig. \ref{fig:LC_states}(b), being invariant under a $m_y$ or $m_x$  mirror reflection combined with time reversal. Moreover, Fig. \ref{fig:bands}(d) has twofold rather than sixfold rotational symmetry, which is a manifestation of the nematic order induced in the AM phase.

\section*{Discussion}
In summary, we showed that altermagnetic states can also be realized in compensated collinear magnets with an odd number of subattices, provided the magnetic moment is not forced to be uniform. We demonstrated this mechanism for a kagome lattice that undergoes intertwined CDW and LC instabilities, in which case the non-uniform magnetic moments are generated by the loop-current patterns. A natural material candidate to realize this phenomenon are the kagome metals $A$V$_3$Sb$_5$, which display vHs near the Fermi level \cite{Kang2022}. Theoretically, different microscopic models have found interaction regimes that favor a LC state \cite{Denner2021,Tazai2022,Ferrari2022,Ziqiang2023,Scammell2023,Wu2023,HYKee2024,fu2025exotic,chen2024flux}. Experimentally, different probes have reported evidence for TRSB either coincident or inside the CDW phase, consistent with a loop-current state \cite{Jiang2021,mielke2022time,khasanov2022time,guguchia2023tunable,bonfa2024unveiling,xing2024optical,guo2022switchable,yu2021evidence,asaba2024evidence,Gui2025}. On the other hand, other experiments report no evidence of TRSB \cite{saykin2023high,farhang2023unconventional,li2022no_observation,koshelev2024}. It thus remains to be established whether the reported signatures of TRSB are due to a spontaneous broken-symmetry phase or induced by weak external strain or magnetic fields. For this reason, additional experiments that can unambiguously assess the existence of loop-current order are desirable.
In this regard, it is interesting to note that our model for a $d$-wave kagome orbital-AM is the 2D version of one of the LC candidates for  $A$V$_3$Sb$_5$. This LC configuration, dubbed congruent CDW flux phase, was proposed in  Ref. \cite{xing2024optical} to explain the observed piezomagnetic response of the STM-measured CDW peaks to an external magnetic field, as well as the absence of a spontaneous Hall effect \cite{saykin2023high,farhang2023unconventional} and the reported breaking of threefold rotational symmetry \cite{Xu2022Three-state,Li2022}. Indeed, piezomagnetism with zero net magnetization is a hallmark of $d$-wave altermagnets \cite{Steward2023,Mcclarty2024}. Our results for the  $d$-wave spin-splitting of the electronic bands provide a compelling way not only to unambiguously identify the realization of this state in   $A$V$_3$Sb$_5$, but also to quantitatively extract the elusive LC order parameter. Given the good energy resolution of previous ARPES experiments in these compounds \cite{Kang2022,Kang2023},  the predicted spin-splittings should be detectable via spin-resolved ARPES. As we show in the SM, these spin-splittings are directly proportional to the magnitude of the LC order parameter and of the SOC. In summary, our work not only significantly expands the types of lattices that can display altermagnetic-like states, but also provides important insights into the unconventional CDW phase of $A$V$_3$Sb$_5$.
\begin{acknowledgments}
We thank M. Christensen, G. Palle, and E. Ritz for fruitful discussions. A.R.C. and R.M.F. were supported by the Air Force Office of Scientific Research under Award No. FA9550-21-1-0423. F.Y. and T.B. were supported by the NSF CAREER grant DMR-2046020.
\end{acknowledgments}

\section*{Data Availability}
The data that support the findings of this article are openly available at \cite{code}.

\bibliography{AMpaper_references}

\end{document}

\title{Supplemental Material: Orbital altermagnetism on the kagome lattice and possible application to $A$V$_3$Sb$_5$}
\author{Anzumaan R. Chakraborty}
\affiliation{Department of Physics, The Grainger College of Engineering, University of Illinois Urbana-Champaign, Urbana, IL 61801, USA}
\affiliation{Anthony J. Leggett Institute for Condensed Matter Theory, The Grainger College of Engineering, University of Illinois Urbana-Champaign, Urbana, 61801, IL, USA}

\author{Fan Yang}
\affiliation{Department of Chemical Engineering and Materials Science,
University of Minnesota, Minneapolis, Minnesota 55455, USA}

\author{Turan Birol}
\affiliation{Department of Chemical Engineering and Materials Science,
University of Minnesota, Minneapolis, Minnesota 55455, USA}

\author{Rafael M. Fernandes}
\affiliation{Department of Physics, The Grainger College of Engineering, University of Illinois Urbana-Champaign, Urbana, IL 61801, USA}
\affiliation{Anthony J. Leggett Institute for Condensed Matter Theory, The Grainger College of Engineering, University of Illinois Urbana-Champaign, Urbana, 61801, IL, USA}

\date{\today}
\maketitle
\tableofcontents
\section{Details of the free energy analysis}
\subsection{Landau expansion in terms of CDW-LC order parameters}

In this supplementary section, we analyze phenomenologically the various magnetic phases that emerge from the coupled CDW order parameter $\textbf{W}=(W_1,W_2,W_3)$ and the LC order parameter  $\boldsymbol{\Phi}=(\Phi_1,\Phi_2,\Phi_3)$. As discussed in the main text, $W_i$, $\Phi_i$ have wave-vector $M_i$ and transform as the $M_1^+$ and $mM_2^+$ irreducible representations (irreps) of the parent magnetic space group (MSG) $P6/mmm.1'$, respectively. %Formally, the little point group of the M-point is isomorphic to $D_{2h}$, which admits only one-dimensional irreps, and because the three M-points in the Brillouin zone (BZ) constitute the \textit{star} of M, the irreps $M_1^+$ and $mM_2^+$ are necessarily three-dimensional. 
The coupled Landau free energy was obtained in Ref. \cite{christensen2022toroidal} using group theory. To quartic order, it is given by 
\begin{equation}
\begin{split}
    \mathcal{F}(\textbf{W},\boldsymbol{\Phi})&=\mathcal{F}_\text{cdw}(\textbf{W})+\mathcal{F}_\text{lc}(\boldsymbol{\Phi})+\mathcal{F}_\text{mixed}(\textbf{W},\boldsymbol{\Phi})\\
    \mathcal{F}_\text{cdw}(\textbf{W})&=\frac{a_w}{2}(\textbf{W}\cdot \textbf{W})+\frac{\gamma_w}{3}W_1W_2W_3+\frac{u_w}{4}(\textbf{W}\cdot \textbf{W})^2+\frac{\lambda_w}{4}(W_1^2W_2^2+W_2^2W_3^2+W_3^2W_1^2)\\
    \mathcal{F}_\text{lc}(\boldsymbol{\Phi})&= \frac{a_\Phi}{2}(\boldsymbol{\Phi}\cdot \boldsymbol{\Phi})+\frac{u_\Phi}{4}(\boldsymbol{\Phi}\cdot \boldsymbol{\Phi})^2+\frac{\lambda_\Phi}{4}(\Phi_1^2\Phi_2^2+\Phi_2^2\Phi_3^2+\Phi_3^2\Phi_1^2)\\
    \mathcal{F}_\text{mixed}(\textbf{W},\boldsymbol{\Phi})&= \frac{\gamma}{3}(\Phi_1\Phi_2 W_3 + \Phi_2\Phi_3 W_1 + \Phi_3 \Phi_1 W_2)+\frac{\lambda_1}{4}(W_1 W_2 \Phi_1 \Phi_2 + W_2 W_3 \Phi_2 \Phi_3 + W_3 W_1 \Phi_3 \Phi_1)\\
    &+\frac{\lambda_2}{4}(W_1^2\Phi_1^2 + W_2^2\Phi_2^2 + W_3^2\Phi_3^2)+\frac{\lambda_3}{4}(\textbf{W}\cdot\textbf{W})(\boldsymbol{\Phi}\cdot \boldsymbol{\Phi})
\end{split}
\label{CDW_free_energy}
\end{equation}
where ($a_w$, $\gamma_w$, $u_w$, $\lambda_w$), ($a_\Phi$, $u_\Phi$, $\lambda_\Phi$), and ($\gamma$, $\lambda_1$, $\lambda_2$, $\lambda_3$) are phenomenological constants that couple $\textbf{W}$-only, $\boldsymbol{\Phi}$-only, and $\textbf{W}$-$\boldsymbol{\Phi}$ mixed terms, respectively. We assume that the quartic terms satisfy the necessary conditions for stability of the free-energy. We also assume the forms $a_w=T-T_{W,0}$ and $a_\Phi=T-T_{\Phi,0}$ for the quadratic coefficients, where $T_{W,0}$ and $T_{\Phi,0}$ are the bare CDW and LC transitions. A free energy of the same form as  Eq. (\ref{CDW_free_energy}), but with $\boldsymbol{\Phi}$ replaced by the L-point CDW order parameter, was systematically analyzed in Ref. \cite{christensen2021theory}. Here, we focus on the situation where the CDW order always emerges first, $T_W>T_\Phi$. The subsequent LC-induced magnetic instabilities are analyzed within a Landau theory of composite order parameters in the next section.\\

The CDW instability must be towards a triple-Q phase due to the presence of the cubic term in $\mathcal{F}_\text{cdw}$. Such a cubic term is allowed by the time-reversal (TR) invariance of $\textbf{W}$. Clearly, a nonzero cubic coupling $\gamma_w$ favors $W_i\neq0$ for every $i$, with $\sgn \gamma_w = -\sgn W_1W_2W_3$. Consequently, the cubic term raises the transition temperature $T_{W,0}\to T_W$ in addition to rendering it first-order.:

\begin{equation}
    T_W=T_{W,0}+\frac{2\gamma_w^2}{81(3u_w+\lambda_w)}
\end{equation}

The resulting CDW state is fourfold degenerate, corresponding to (for $W>0$) the configurations  ($W,W,W$), ($-W,-W,W$), ($W,-W,-W$), ($-W,W,-W$) for $\gamma_w<0$, corresponding to the trihexagonal phase; and ($-W,-W,-W$), ($-W,W,W$), ($W,-W,W$), ($W,W,-W$) for $\gamma_w>0$, corresponding to the Star-of-David phase. We stress that while these two phases are non-degenerate when $\gamma_w \neq 0$ and correspond to different dimerization patterns, they yield the same space group. Specifically, their condensation breaks translation symmetry but preserves all point group symmetries, thus resulting in a space group that is isomorphic to the parent space group (this is possible because they are both infinite groups). Thus, unless otherwise stated in this section, we refer to both as triple-Q CDW phases and denote them as $(W,W,W)$. A relatively simple epxression for $W$ follows by minimizing $\mathcal{F}_\text{cdw}$ for $\boldsymbol{\Phi}=0$:
\begin{equation}
    |W|=\frac{|\gamma_w|+\sqrt{\gamma_w^2-36a_w(3u_w+\lambda_w)}}{4(3u_w+\lambda_w)}
    \label{W_in_cdw}
\end{equation}

While the leading instability of the CDW-only problem is always toward the triple-Q CDW phase, other CDW patterns may emerge due to the subsequent condensation of the LC order parameter $\boldsymbol{\Phi}$ at a lower temperature $T_\Phi<T_W$. These mixed phases are rather complex owing to the mixed terms in Eq. (\ref{CDW_free_energy}). In Table \ref{tab:GS_table}, we show all symmetry-distinct minima of coexistence between the CDW and LC order parameters allowed by group theory, as obtained using the Isosubgroup program \cite{stokes2016isosubgroup,isosubgroup}. The corresponding order parameter configurations and resulting magnetic space groups are also given.

The phase diagram in the main text was obtained by comparing the free-energy minima corresponding to phases 0, 1, 2, 3, and 4. The Landau parameters used were the same as in Ref. \cite{christensen2021theory}, which, as we explained above, performed a systematic numerical analysis of the free energy (\ref{CDW_free_energy}). The parameters were: $\gamma_w=0.25$, $\lambda_w=0.6$, $u_w=1.2$, $\lambda_\Phi=0.8$, $u_\Phi=1.5$, $\lambda_1=0.01$, $\lambda_2=0.015$, $\lambda_3=0.0175$, $T_{W,0}=0$, and $T_{\Phi,0}=-0.02$.

\begin{table*}[t]
\centering
\footnotesize
\renewcommand{\arraystretch}{1.2}

\begin{tabular}{|c|c|c|c|c|c|}
\hline
phase & OP configuration & magnetic subgroup & TR-odd multipoles & TR-even multipoles & properties\\
\hline\hline
0 & $(W,W,W)$ & $P6/mmm.1'$ & - & - & 3Q CDW\\
\hline\hline
1 & \parbox[c]{2cm}{\centering $(W,W,W)$\\$(\Phi,\Phi,\Phi)$} &
    $P6/mm'm'$ &
    $\mathcal{M}=W\Phi$ & - & ferromagnetic\\
\hline
2 & \parbox[c]{2cm}{\centering $(W,0,0)$\\$(0,\Phi,\Phi)$} &
    $Cmmm.1_a'$ &
    - &
    \parbox[c]{3.5cm}{\centering $\boldsymbol{\mathcal{Q}}=\Phi^2\langle-1,\sqrt{3}\rangle$} &
    \parbox[c]{3.5cm}{\centering antiferromagnetic\\nematic}\\
\hline
3 & \parbox[c]{2cm}{\centering $(W,W',W)$\\$(\Phi,0,-\Phi)$} &
    $Cmmm.1$ &
    \parbox[c]{4cm}{\centering $\boldsymbol{\mathcal{N}}=W\Phi\langle 1,0\rangle$\\$\Upsilon=-W\Phi^3$} &
    \parbox[c]{3.5cm}{\centering $\boldsymbol{\mathcal{Q}}=\Phi^2\langle1,0\rangle$} &
    \parbox[c]{3.5cm}{\centering $d$-wave altermagnetic\\$i$-wave altermagnetic\\nematic}\\
\hline
4 & \parbox[c]{2cm}{\centering $(W,W',W)$\\$(\Phi,\Phi',\Phi)$} &
    $Cm'm'm$ &
    \parbox[c]{4cm}{\centering $\boldsymbol{\mathcal{N}}=(W\Phi-W'\Phi')\langle0,1\rangle$\\$\mathcal{M}=2W\Phi+W'\Phi'$} &
    \parbox[c]{3.5cm}{\centering $\boldsymbol{\mathcal{Q}}=(\Phi^2-\Phi'^2)\langle1,0\rangle$} &
    \parbox[c]{3.5cm}{\centering $d$-wave altermagnetic\\ferromagnetic\\nematic}\\
\hline
5 & \parbox[c]{2cm}{\centering $(W,0,0)$\\$(0,\Phi,\Phi')$} &
    $P2/m.1'_a$ &
    - &
    \parbox[c]{3.5cm}{\centering $\boldsymbol{\mathcal{Q}}=\langle\Phi'^2-2\Phi^2,\sqrt{3}\Phi'^2\rangle$\\$\mathcal{G}=\Phi^2\Phi'^2(\Phi^2-\Phi'^2)$} &
    \parbox[c]{3.5cm}{\centering antiferromagnetic\\nematic\\ferroaxial}\\
\hline
6 & \parbox[c]{2cm}{\centering $(W,0,0)$\\$(\Phi,0,0)$} &
    $Pm'm'm$ &
    \parbox[c]{4cm}{\centering $\mathcal{M}=W\Phi$\\$\boldsymbol{\mathcal{N}}=W\Phi\langle\sqrt{3},1\rangle$} &
    \parbox[c]{3.5cm}{\centering $\boldsymbol{\mathcal{Q}}=\Phi^2\langle1,-\sqrt{3}\rangle$} &
    \parbox[c]{3.5cm}{\centering ferromagnetic\\$d$-wave altermagnetic\\nematic}\\
\hline
7 & \parbox[c]{2cm}{\centering $(W,W',W'')$\\$(\Phi,\Phi',\Phi'')$} &
    $P2/m.1$ &
    all & all & all\\
\hline
\end{tabular}

\caption{Symmetry-distinct mixed CDW-LC phases resulting from the combination of a CDW order parameter $\textbf{W}$ that transforms as the $M_1^+$ irrep and an LC order parameter $\boldsymbol{\Phi}$ that transforms as the $mM_2^+$ irrep. Row $0$ describes the ``pure'' triple-Q CDW phase, and rows $1$–$7$ describe the phase along with their order parameter CDW-LC configuration, magnetic space group, nonzero time-reversal (TR)-odd $\textbf{Q}=0$ multipoles, non-zero TR-even $\textbf{Q}=0$ multipoles, and corresponding physical properties.}
\label{tab:GS_table}
\end{table*}

\subsection{Composite order parameters}

By combining  $\textbf{W}$ and  $\boldsymbol{\Phi}$, we can construct $\textbf{Q}=0$ composites, i.e., polynomials of $W_i$ and $\Phi_i$ that transform as $\Gamma$-point irreps of the parent MSG $P6/mmm.1'$. Establishing which composites are non-zero for each phase listed in Table \ref{tab:GS_table} sheds further light on their properties. Note that, because $\Phi_i\to-\Phi_i$ under TR, it follows that TR-even/non-magnetic (TR-odd/magnetic) composites must be even (odd) in $\Phi_i$. With this in mind, it is convenient to focus on the properties of the irrep $mM_2^+$ under spatial symmetries alone, which is the same as focusing on the properties of the irrep  $M_2^+$. 
\begin{table*}[]
    \centering
    \renewcommand{\arraystretch}{1.2} 
\begin{tabular}{ c|| c | c | c | c | c |}
$\otimes$&$\Gamma_1^+$&$\Gamma_2^+$&$\Gamma_5^+$&$M_1^+$&$M_2^+$ \\
\hline
\hline
$\Gamma_1^+$&$\Gamma_1^+$&-&-&-&- \\
\hline
$\Gamma_2^+$&$\Gamma_2^+$&$\Gamma_1^+$&-&-&- \\
\hline
$\Gamma_5^+$&$\Gamma_5^+$&$\Gamma_5^+$&$\Gamma_1^+\oplus \Gamma_2^+\oplus \Gamma_5^+$&-&- \\
\hline
$M_1^+$&$M_1^+$&$M_2^+$&$M_1^+\oplus M_2^+$&$M_1^+\oplus M_2^+\oplus \Gamma_1^+\oplus \Gamma_5^+$&- \\
\hline
$M_2^+$&$M_2^+$&$M_1^+$&$M_1^+\oplus M_2^+$&$M_1^+\oplus M_2^+\oplus \Gamma_2^+\oplus \Gamma_5^+$&$M_1^+\oplus M_2^+\oplus \Gamma_1^+\oplus \Gamma_5^+$ \\
\hline
\end{tabular}
    \caption{Irrep product decomposition table of a collection of $P6/mmm$ irreps $\mathcal{C}=\{\Gamma_1^+,\Gamma_2^+,\Gamma_5^+,M_1^+,M_2^+\}$. Since the table is self-contained, arbitrary products of $M_1^+$ and $M_2^+$ harmonics can only generate themselves and the other $\Gamma$-point irreps in $\mathcal{C}$. Thus, including TR, the five $\Gamma$-point irreps $m\Gamma_2^+$, $m\Gamma_5^+$, $m\Gamma_1^+$, $\Gamma_5^+$, $\Gamma_2^+$ exhaust all symmetry-breaking $\textbf{Q}=0$ composites that can be generated by $\textbf{W}$ and $\boldsymbol{\Phi}$.}
    \label{tab:irrep_closure}
\end{table*}

To construct the composites, it is helpful to note that the set of irreps $\mathcal{C}=\{ \Gamma_1^+,\Gamma_2^+,\Gamma_5^+,M_1^+,M_2^+\}$ are closed under direct products $\otimes$ and direct sums $\oplus$. This closure property can be verified explicitly from the irrep product decompositions given in Table \ref{tab:irrep_closure}, and follows from the character table of the $M_i^\pm$ irreps. Consequently, the following five $\Gamma$-point irreps exhaust all symmetry-breaking $\textbf{Q}=0$ composites that can be generated by combinations of $\textbf{W}$ and $\boldsymbol{\Phi}$: $m\Gamma_2^+=A_{2g}^-$, $m\Gamma_5^+=E_{2g}^-$, $m\Gamma_1^+=A_{1g}^-$, $\Gamma_5^+=E_{2g}^+$, $\Gamma_2^+=A_{2g}^+$. In these expressions, we also included the equivalent Mulliken symbol, using the superscript $\pm$ to denote even/odd under TR. To lowest order in the order parameters and to linear order in $\textbf{W}$, the TR-odd composites are given by:
\begin{align}
    m\Gamma_2^+&=A_{2g}^-\text{: }\quad\mathcal{M}=\textbf{W}\cdot\boldsymbol{\Phi}\\
    m\Gamma_5^+&=E_{2g}^-\text{: }\quad\boldsymbol{\mathcal{N}}=\begin{pmatrix}
        \sqrt{3}(W_1\Phi_1-W_3\Phi_3) \\ W_1\Phi_1 + W_3\Phi_3-2W_2\Phi_2
    \end{pmatrix}=\mathcal{N}\begin{pmatrix}
        \cos\theta\\\sin\theta
    \end{pmatrix}\label{N_def}\\
    m\Gamma_1^+&=A_{1g}^-\text{: }\quad\Upsilon=W_1\Phi_1(\Phi_2^2-\Phi_3^2)+W_2\Phi_2(\Phi_3^2-\Phi_1^2)+W_3\Phi_3(\Phi_1^2-\Phi_2^2).
\end{align}
The irreps allow us to immediately identify the physical meaning of each composite (see, for instance, \cite{Fernandes2024_AM}). $\mathcal{M}$ is a FM (ferromagnetic) order parameter, whereas  $\boldsymbol{\mathcal{N}}$ and $\Upsilon$ are AM (altermagnetic) order parameters with $d$-wave and $i$-wave symmetries, respectively. In all three cases, the magnetic moments point out-of-plane (i.e., along the $z$-axis). For later convenience, the angle $\theta$ is introduced to cast $\boldsymbol{\mathcal{N}}$ in polar form. These results can be verified by combining polynomials of the in-plane momentum  $\textbf{k}=k(\cos\theta_k,\sin\theta_k)$, which transforms as  $m\Gamma_6^- = E_{1u}^{-}$, and the out-of-plane spin polarization $\sigma_z$, which transforms as the $m\Gamma_2^+=A_{2g}^-$ irrep: 
\begin{align}
    \mathcal{M}&\sim\sigma_z\\
    \boldsymbol{\mathcal{N}}&\sim \sigma_zk^2\begin{pmatrix}
        \sin2\theta_k\\\cos2\theta_k
    \end{pmatrix}\\
    \Upsilon&\sim \sigma_zk^6\sin6\theta_k
\end{align}
On the other hand, the TR-even composites are given by
\begin{align}
    \Gamma_5^+&=E_{2g}^+\text{: }\quad\boldsymbol{\mathcal{Q}}=\begin{pmatrix}
        \Phi_1^2+\Phi_3^2-2\Phi_2^2\\\sqrt{3}(\Phi_3^2-\Phi_1^2)
    \end{pmatrix}=\mathcal{Q}\begin{pmatrix}
        \cos\phi\\\sin\phi
    \end{pmatrix}\\
    \Gamma_2^+&=A_{2g}^+\text{: }\quad\mathcal{G}=(\Phi_1^2-\Phi_2^2)(\Phi_2^2-\Phi_3^2)(\Phi_3^2-\Phi_1^2)
\end{align}
As before, based on the symmetry properties of the irreps, we identify $\boldsymbol{\mathcal{Q}}$ as a nematic order parameter (electric quadrupole moment), whose condensation breaks three-fold rotation symmetry, and $\mathcal{G}$ as a ferroaxial order parameter (electric toroidal dipole moment), whose condensation breaks all vertical mirrors \cite{DayRoberts2025}. Using the definitions of the TR-even and TR-odd composites, it is straightforward to determine which  $\textbf{Q}=0$ composites are non-zero in each phase. The results are shown in the last column of Table \ref{tab:GS_table}.

It is interesting to note the connection between the nematic $\boldsymbol{\mathcal{Q}}$ and the $d$-wave AM order parameter $\boldsymbol{\mathcal{N}}=\mathcal{N}(\cos\theta,\sin\theta)$. A real-space $C_{3z}$ rotation, which is equivalent to an M-point cyclic permutation, amounts to a three-fold rotation of both order parameter angles: $\theta\to\theta+\frac{2\pi}{3}$,  $\phi\to\phi+\frac{2\pi}{3}$. Consequently, in addition to breaking TR, the crystalline symmetries broken by $\boldsymbol{\mathcal{N}}$ are precisely the same as those broken by $\boldsymbol{\mathcal{Q}}$. Thus, condensation of $\boldsymbol{\mathcal{N}}$ implies condensation of $\boldsymbol{\mathcal{Q}}$, and the six-fold rotation symmetry of the CDW-ordered crystal is broken down to two-fold via nematicity upon $d$-wave AM order. Indeed, this can also be seen directly from the fact that $\Gamma_5^+\in m\Gamma_5^+\otimes m\Gamma_5^+$. Thus, when the $d$-wave AM order parameter is nonzero, it induces a nematic distortion according to $\mathcal{Q}_1 \sim \mathcal{N}_1^2 - \mathcal{N}_2^2$ and $\mathcal{Q}_2 \sim -2 \mathcal{N}_1 \mathcal{N}_2$, or, equivalently:
\begin{equation}
\mathcal{Q} \sim \mathcal{N}^2 \, \,\mathrm{and} \, \,  \phi = -2\theta \label{eq:Qind} 
\end{equation}

\begin{figure*}
\includegraphics[width=1.0\columnwidth]{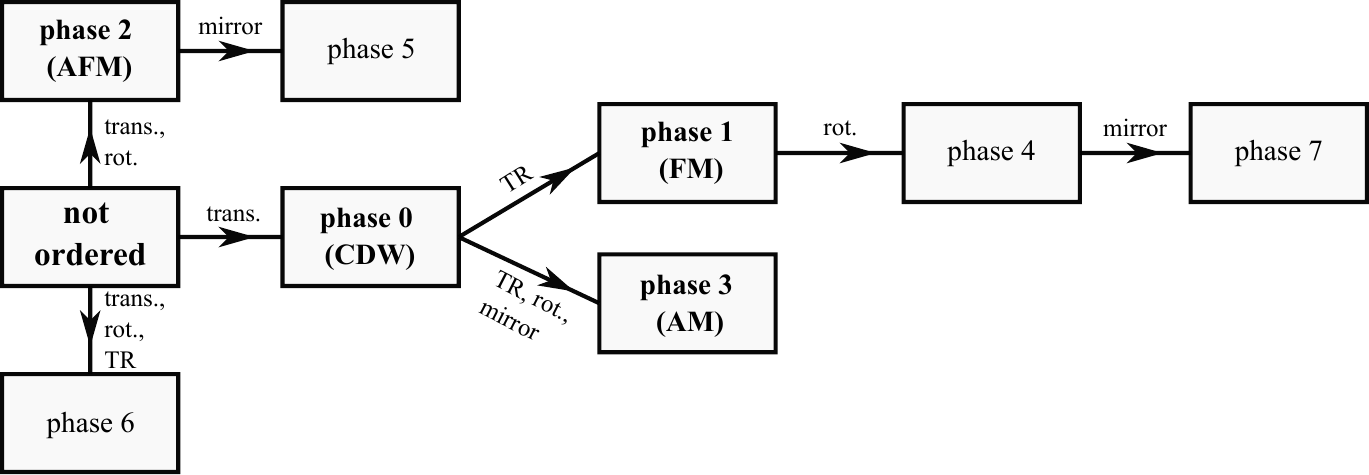}\caption{An exhaustive diagram of subgroup relations between the symmetry groups of the different phases obtained by minimizing $\mathcal{F}(\textbf{W},\boldsymbol{\Phi})$, where a path of one or more arrows linking phase A to phase B implies that the latter can be reached from the former by breaking the listed symmetries. The symmetries that are broken, namely, translations (trans.), rotations (rot.), time-reversal (TR), and mirror are listed next to the corresponding arrow. The bold phases are studied in the main text.}\label{fig:subgroups}
\end{figure*}

We emphasize that the discrete symmetry of the crystal implies that there are infinitely many higher-order polynomials of $\textbf{W}$ and $\boldsymbol{\Phi}$ that transform as each of these irreps. As it will be important in Section II of this Supplementary Material, we list the first higher-order $A_{2g}^-$ ferromagnetic composites $\mathcal{M}_1$ and $\mathcal{M}_2$: 
\begin{align}
    \mathcal{M}_1&=\Phi_1\Phi_2\Phi_3\\
    \mathcal{M}_2&=\Phi_1W_2W_3+\Phi_2W_3W_1+\Phi_3W_1W_2. \label{eq:M12}
\end{align}
Crucially, the composites are obviously not independent owing to the closure of the collection of irreps. In particular, combinations of one or more multipoles can induce other multipoles, as can be inferred directly from the irrep product decompositions in Table \ref{tab:irrep_closure}. Consider for example that $m\Gamma_2^+,m\Gamma_1^+\in \Gamma_5^+\otimes m\Gamma_5^+$ and $\Gamma_2^+\in \Gamma_5^+\otimes \Gamma_5^+\otimes \Gamma_5^+$. This implies that the ``composites of composites'' $\mathcal{N}_1\mathcal{Q}_2-\mathcal{N}_2\mathcal{Q}_1$, $\boldsymbol{\mathcal{N}}\cdot \boldsymbol{\mathcal{Q}}$, and $3\mathcal{Q}_1^2\mathcal{Q}_2-\mathcal{Q}_2^3$ transform as a FM, $i$-wave AM, and a ferroaxial order parameter, respectively. Indeed, the latter two coincide with our definitions of $\Upsilon$ and $\mathcal{G}$ in terms of the CDW-LC order parameters (up to an arbitrary pre-factor):
\begin{align}
\Upsilon&=\boldsymbol{\mathcal{N}}\cdot\boldsymbol{\mathcal{Q}}=\mathcal{N}\mathcal{Q}\cos(\theta-\phi) \sim \mathcal{N}^3 \cos 3\theta \\
\mathcal{G}&=3\mathcal{Q}_1^2\mathcal{Q}_2-\mathcal{Q}_2^3=\mathcal{Q}^3\sin3\phi\sim \mathcal{N}^6 \sin{6 \theta}.\label{eq:Upsilon}
\end{align}
where, in the last equations, we used the result that  $\phi = -2\theta$ and $\mathcal{Q} \sim \mathcal{N}^2$ from Eq. (\ref{eq:Qind}).  As for the first ``composite of composites,'' it simply defines a new $A_{2g}^-$ FM composite:
\begin{align}
    \mathcal{M}_3&=\mathcal{N}_1\mathcal{Q}_2-\mathcal{N}_2\mathcal{Q}_1=\mathcal{N}\mathcal{Q}\sin(\theta-\phi) \sim \mathcal{N}^3 \sin{3 \theta}\\&=W_1\Phi_1(\Phi_2^2+\Phi_3^2-2\Phi_1^2)+W_2\Phi_2(\Phi_3^2+\Phi_1^2-2\Phi_2^2)+W_3\Phi_3(\Phi_2^2+\Phi_2^2-2\Phi_3^2). \label{eq:M3}
\end{align}
where we used, once again, Eq. (\ref{eq:Qind}). The key point, which we confirm in the following subsection via mean field theory is that the relative orientation of the condensed nematic and $d$-wave AM composites can induce either FM or $i$-wave AM that are both cubic in $\boldsymbol{\Phi}$ and thus parametrically weaker than either $\mathcal{N}\propto|\boldsymbol{\Phi}|$, and $\mathcal{Q}\propto \mathcal{N}^2\propto |\boldsymbol{\Phi}|^2$.

\subsection{Landau expansion in terms of composites: leading-order terms}
Our interest is to analyze the emergence and interplay of the LC-induced $\textbf{Q}=0$ multipoles (FM and AM) from within the CDW state. Thus, it is instructive to directly cast $\mathcal{F}(\textbf{W},\boldsymbol{\Phi})$ in terms of the composites developed in the previous subsection along with setting $\textbf{W}=(W,W,W)$. In this case, the leading composites for nonzero LC order are $\mathcal{M}$ and $\boldsymbol{\mathcal{N}}$, which are both linear in $\Phi_i$. Thus, for simplicity, we treat these as primary orders, whose condensation may induce secondary multipoles. It is important to recognize that near $T_\Phi$, where $W$ is already finite, the most singular degree of freedom is $\boldsymbol{\Phi}$ as its emergence renders $\mathcal{M}$ and/or $\boldsymbol{\mathcal{N}}$ nonzero. With this in mind, it is appropriate to treat $W$ as a constant (i.e., a frozen degree-of-freedom) that can be absorbed into the Landau coefficients. In doing so, $\mathcal{M}$ and $\boldsymbol{\mathcal{N}}$ appear as linear functions of the critical degrees-of-freedom according to:
\begin{align}
\mathcal{M}&=W(\Phi_1+\Phi_2+\Phi_3)
\label{M_simplified}\\
\boldsymbol{\mathcal{N}}&=W\begin{pmatrix}
        \sqrt{3}(\Phi_1-\Phi_3) \\ \Phi_1 + \Phi_3-2W_2\Phi_2
    \end{pmatrix}.
\label{N_simplified}
\end{align}
Upon inverting, we obtain 
\begin{align}
\Phi_1&=\frac{2\mathcal{M}+\sqrt{3}\mathcal{N}_1+\mathcal{N}_2}{6W}\\
\Phi_2&=\frac{\mathcal{M}-\mathcal{N}_2}{3W}\\
\Phi_3&=\frac{2\mathcal{M}-\sqrt{3}\mathcal{N}_1+\mathcal{N}_2}{6W}
\end{align}
Substituting it in Eq. (\ref{CDW_free_energy}) yields a new Landau expansion $\mathcal{F}(\textbf{W},\boldsymbol{\Phi})\to F(\mathcal{M},\boldsymbol{\mathcal{N}})$ given by 
\begin{equation}
    F(\mathcal{M},\boldsymbol{\mathcal{N}}) = \frac{a_1}{2}\mathcal{M}^2+\frac{a_2}{2}\mathcal{N}^2+\frac{u_1}{4}\mathcal{M}^4 \\
     +\frac{u_2}{4}\mathcal{N}^4+\frac{\zeta}{6}\mathcal{N}^6\cos6\theta \\
     +\frac{u}{4} \mathcal{M}^2\mathcal{N}^2+\frac{\lambda}{4}\mathcal{M}\mathcal{N}^3\sin 3\theta
     \label{simplified_FE}
\end{equation}
with new Landau parameters
\begin{equation}
    a_1=\frac{T-T_\mathcal{M}}{3W^2}\text{, }\quad a_2=\frac{T-T_{\mathcal{N}}}{6W^2}\text{, }\quad u_1=\frac{3u_\Phi+\lambda_\Phi}{27W^4}\text{, }\quad u_2=\frac{4u_\Phi+\lambda_\Phi}{144W^4}\text{, }\quad u=\frac{u_\Phi}{9}\text{, and }\quad \lambda=-\frac{\lambda_\Phi}{54}
\end{equation}
and temperature scales
\begin{align}
    T_\mathcal{M}&\equiv T_{\Phi,0}-\frac{2W\gamma}{3}-\frac{\lambda_1+\lambda_2+3\lambda_3}{2}W^2
    \label{TM}\\
    T_{\mathcal{N}}&\equiv T_{\Phi,0}+\frac{W\gamma}{3}-\frac{-\lambda_1+2\lambda_2+6\lambda_3}{4}W^2.
    \label{TN}
\end{align}
We note that the symmetry-allowed sixth-order coefficient $\zeta$ does not appear to this order in the expansion in $W$. As we show in Section II-D, it emerges as higher-order terms are included.

We stress that $T_\mathcal{M}$ ($T_{\mathcal{N}}$), which is renormalized by the mixed terms in $\mathcal{F}$, only refers to the FM (AM) transition when $T_\mathcal{M}>T_{\mathcal{N}}$ ($T_\mathcal{M}<T_{\mathcal{N}}$). This is because the condensation of one order parameter strongly renormalizes the quadratic coefficient of the other through anharmonic mixed terms, e.g., $u\mathcal{M}^2\mathcal{N}^2$. With this in mind, to $\mathcal{O}(W)$ the orbital magnetic LC instability leads to FM (AM) order when $W\gamma<0$ ($W\gamma>0$). The expressions quoted in the main text are recovered by considering $W<0$. For large $W$, the phase that is realized depends on the relative sign and magnitude of the $W^2$ coefficients. By comparing the expressions for $T_\mathcal{M}$ and $T_{\mathcal{N}}$, it follows that $T_\mathcal{M}>T_{\mathcal{N}}$ ($T_{\mathcal{N}}>T_\mathcal{M}$) when $W\gamma<0$ ($W\gamma>0$) provided that 
\begin{equation}
    |W|<\frac{4|\gamma|}{|\lambda_1+4\lambda_2+12\lambda_3|}.
\end{equation}
Otherwise, for either sign of $W$, $T_\mathcal{M}>T_{\mathcal{N}}$ ($T_{\mathcal{N}}>T_\mathcal{M}$) when $\lambda_1+4\lambda_2+12\lambda_3<0$ ($\lambda_1+4\lambda_2+12\lambda_3>0$).

We now provide a qualitative description of the phases hosted by this composite Landau theory. We first consider the case of $T_\mathcal{M}>T_{\mathcal{N}}$, which is rather straightforward. In this scenario, a second-order FM transition occurs at $T=T_\mathcal{M}$ with $\mathcal{M}$ acquiring a non-zero value. Due to the coupling of $\mathcal{M}$ to $\mathcal{N}^3$, as long as $\lambda\mathcal{M}$ is small enough, $\boldsymbol{\mathcal{N}}=0$ and only $D_{6h}$-preserving FM arises. The corresponding state in Table \ref{tab:GS_table} is phase 1.

The case of $T_{\mathcal{N}}>T_\mathcal{M}$ is richer. At $T=T_{\mathcal{N}}$ the two-component $d$-wave AM order $\boldsymbol{\mathcal{N}}=\mathcal{N}(\cos\theta,\sin\theta)$ condenses. In this case, the sign of $\zeta$ determines $\theta$, whose value signals the presence of certain secondary multipoles. When $\zeta>0$, the sixth-order term favors $\theta=(2n+1)\frac{\pi}{6}$, with $n=0,1,...,5$. In such a scenario, $|\sin 3\theta|=1$ and an effective conjugate field $H_\text{eff}=\lambda\mathcal{N}^3$ couples to $\mathcal{M}$, thus inducing a weak FM moment that is proportional to $\mathcal{N}^3$. This result can also be understood from Eq. (\ref{eq:M3}), since $\mathcal{M}_3$ is non-zero for these values of $\theta$. The corresponding state in Table \ref{tab:GS_table} is phase 4. On the other hand, when $\zeta<0$, the angles $\theta=n\frac{\pi}{3}$ are favored. As a result, $\mathcal{M}_3$ in Eq. (\ref{eq:M3}) becomes zero, indicating the absence of FM order, while $\Upsilon$ in Eq. (\ref{eq:Upsilon}) becomes non-zero, indicating that an $i$-wave AM order proportional to $\mathcal{N}^3$ is induced. The corresponding state in Table \ref{tab:GS_table} is phase 3. 

For completeness, we provide in Fig. \ref{fig:subgroups} an exhaustive diagram of symmetry-breaking pathways that can arise from condensation of one or more composites. Depending on the parameters of $\mathcal{F}$, the symmetry breakings can occur simultaneously or successively, where transitions between phases whose symmetries do not constitute a group-subgroup pair are restricted to be first-order. The particular case studied in the main text is the `not ordered' to phase $0$ (CDW) transition, followed by either a second-order transition into phase $1$ (FM) or phase $3$ (AM), or a first-order transition into phase $2$ (AFM). Note that the FM phase can further break $C_{3z}$ upon condensation of $\boldsymbol{\mathcal{Q}}$, which necessarily triggers an accompanying AM moment $\boldsymbol{\mathcal{N}}$. While phase $4$ has the same symmetry as the AM + ``weak'' FM phase discussed previously, there is no assumed condition on the relative strength between the two order parameters when the phase emerges from the rotation-preserving ferromagnetic phase (phase $1$). Further mirror symmetries can be broken to condense the remaining two composites -- $\mathcal{G}$ and $\Upsilon$ -- corresponding to phase $7$, the lowest symmetry phase possible from these CDW and LC order parameters.

\subsection{Landau expansion in terms of composites: higher-order terms}

As explained above, the expression for the composite Landau free energy $F(\mathcal{M},\boldsymbol{\mathcal{N}} )$ does not show explicitly the sixth-order term with coefficient $\zeta$ that selects between the AM + FM state (phase 4) or the pure AM state (phase 3). Minimization over $\mathcal{M}$, however, yields an effective theory in terms of  $\boldsymbol{\mathcal{N}}=\mathcal{N}(\cos\theta,\sin\theta)$  alone that contains the sixth-order term:
\begin{equation}
    F(\boldsymbol{\mathcal{N}})=\frac{a_2}{2}\mathcal{N}^2 + \frac{u_2}{4}\mathcal{N}^4+\frac{\lambda^2}{64a_1}\mathcal{N}^6\cos 6\theta+...
    \label{N_only_FE_simplified}
\end{equation}
Since $a_1>0$, the sixth-order term always favors $\theta=(2n+1)\pi/6$, and thus gives $\sin3\theta\neq0$, which yields $\mathcal{M}\neq 0$ (phase 4). This is not unexpected, since integrating out the ferromagnetic fluctuations tend to favor a nonzero magnetization. However, this analysis is clearly incomplete, since the explicit minimization of the original free energy shown in the main text results in a wide parameter regime where the pure AM state (phase 3) is realized. This issue is a consequence of freezing out fluctuations in $\textbf{W}$ when expressing the composites in terms of the original CDW and LC order parameters in Eqs. (\ref{M_simplified}) and (\ref{N_simplified}). In other words, the mapping $(\textbf{W},\boldsymbol{\Phi})\to (\mathcal{M},\boldsymbol{\mathcal{N}})$ is not one-to-one.

To remedy this shortcoming, it is necessary to include non-magnetic degrees of freedom explicitly into our analysis. Specifically, we introduce the $E_{2g}^+$ nematic composite $\mathcal{Q}$ and an $A_{1g}^+$ symmetry-preserving composite $\mathcal{V}$. For our purposes, it is convenient to write these composites in terms of the CDW order parameters alone:
\begin{align}
    \boldsymbol{\mathcal{Q}}&=
    \begin{pmatrix} W_1^2+W_3^2-2W_2^2 \\ \sqrt{3}(W_3^2-W_1^2)\end{pmatrix}\\
    \mathcal{V}&= \textbf{W}\cdot\textbf{W}-3W^2.
\end{align}
Near the magnetic transition, $W_i\approx W$; we approximate the composites akin to Eqs. (\ref{M_simplified})-(\ref{N_simplified}) as
\begin{align}
    \boldsymbol{\mathcal{Q}}&=
    W\begin{pmatrix} W_1+W_3-2W_2 \\ \sqrt{3}(W_3-W_1)\end{pmatrix}\\
    \mathcal{V}&= W(W_1+W_2+W_3)-3W^2
\end{align}
where, as before, $W$ takes the known value inside the CDW state (Eq. (\ref{W_in_cdw})). Note that we have introduced a constant shift of $-3W^2$ in $\mathcal{V}$, which ensures that the $\mathcal{V}$ measures the deviation of the CDW fluctuations away from the equal-$W$ triple-$Q$ phase. As a result, $W$ is zero in the CDW phase and becomes non-zero below the LC transition due to the feedback effect of the condensation of $\boldsymbol{\Phi}$ on $\boldsymbol{W}$. Introducing these new composites ensures that $(\textbf{W},\boldsymbol{\Phi})\to (\mathcal{M},\boldsymbol{\mathcal{N}},\boldsymbol{\mathcal{Q}},\mathcal{V})$ constitutes an invertible linear map, and consequently, no CDW degrees of freedom are frozen out. Writing  $\boldsymbol{\mathcal{Q}}=\mathcal{Q}(\cos\phi,\sin\phi)$ and deviating slightly from the parameter notation in Eq. (\ref{simplified_FE}) for clarity, this yields the free energy near the magnetic transition
\begin{equation}
\begin{split}
\mathcal{F}(\mathcal{M},\boldsymbol{\mathcal{N}},\mathcal{V},\boldsymbol{\mathcal{Q}}) &= \frac{\tilde{a}_m}{2}\mathcal{M}^2+\frac{\tilde{a}_n}{2}\mathcal{N}^2+\frac{\tilde{a}_q}{2}\mathcal{Q}^2+\frac{a_v}{2}\mathcal{V}^2+\frac{u_m}{4}\mathcal{M}^4+\frac{u_n}{4}\mathcal{N}^4+\frac{u_q}{4}\mathcal{Q}^4+\frac{u_v}{4}\mathcal{V}^4 \\ &+ \frac{\tilde{\gamma}_q}{3}\mathcal{Q}^3 \cos 3\phi + \frac{\gamma_v}{3}\mathcal{V}^3 + \frac{u_{mn}}{4}\mathcal{M}^2\mathcal{N}^2+\frac{\lambda_{mn}}{4}\mathcal{M}\mathcal{N}^3\sin 3\theta + \frac{\tilde{\gamma}_{nq}}{3}\mathcal{N}^2\mathcal{Q}\cos(2\theta+\phi)\\
&+\frac{\tilde{\gamma}_{mnq}}{3}\mathcal{M}\mathcal{N}\mathcal{Q}\sin(\theta-\phi) + \frac{u_{mq}}{4}\mathcal{M}^2\mathcal{Q}^2 +  \frac{u_{nq}}{4}\mathcal{N}^2 \mathcal{Q}^2 + \frac{u'_{nq}}{4}\mathcal{N}^2\mathcal{Q}^2\cos(2\theta-2\phi)\\ & + \frac{u_{mnq}}{4}\mathcal{M}\mathcal{N}\mathcal{Q}^2 \sin(\theta+2\phi)
\end{split}
\label{simplified_FE_2}
\end{equation}
where the Landau parameters with tildes include anharmonic contributions from $\mathcal{V}$: $\tilde{a}_i=a_i+a_i'\mathcal{V}+a_i''\mathcal{V}^2$ for $i=m,n,q$ and $\tilde{\gamma}_j=\gamma_j+\gamma_j'\mathcal{V}$ for $j=q,nq,mnq$. The quadratic coefficients in $\mathcal{M}$, $\boldsymbol{\mathcal{N}}$, and $\boldsymbol{\mathcal{Q}}$ are
\begin{align}
    a_m&=\frac{T-T_\mathcal{M}}{3W^2},\quad a_m'\frac{2\gamma+3W(\lambda_1 + \lambda_2 + 3\lambda_3)}{27W^3},\quad a_m''=\frac{\lambda_1 + \lambda_2 + 3\lambda_3}{54W^4} \\
    a_n&=\frac{T-T_\mathcal{N}}{6W^2},\quad a_n'=\frac{-2\gamma +3W(-\lambda_1+2\lambda_2+6\lambda_3)}{108W^3},\quad a_n''=\frac{-\lambda_1+2\lambda_2+6\lambda_3}{216W^4}\\
    a_q&=\frac{T-T_{W,0}-\frac{1}{3}W\gamma_w+3W^2u_w}{6W^2},\quad a_q'=\frac{-\gamma_w+18Wu_w}{54W^3},\quad a_q''=\frac{u_w}{18W^4}\\
    a_v&= \frac{T-T_{W,0}+\frac{2}{3}W\gamma_w + 3W^2(\lambda_w+3u_w)}{3W^2}
\end{align}
whereas the quartic coefficients are 
\begin{align}
    u_m &=\frac{3u_\Phi+\lambda_\Phi}{27 W^4},\quad u_n = \frac{4u_\Phi+\lambda_\Phi}{144W^4},\quad u_q=\frac{4u_w+\lambda_w}{144W^4},\quad u_v=\frac{3u_w+\lambda_w}{27W^4} \\
    u_{mn}&=\frac{u_\Phi}{9W^4},\quad  \lambda_{mn}=-\frac{\lambda_\Phi}{54W^4}\\ u_{mq}&=\frac{-\lambda_1+2\lambda_2+6\lambda_3}{108W^4},\quad u_{nq}=\frac{\lambda_1+4\lambda_2+12\lambda_3}{432W^4},\quad u'_{nq}=-\frac{\lambda_1+\lambda_2}{216W^4},\quad
    u_{mnq} = \frac{\lambda_1-2\lambda_2}{108W^4}
\end{align}
and the cubic coefficients are
\begin{align}
    \gamma_q&=\frac{-2\gamma_w+9W\lambda_w}{216W^3}\quad \gamma'_q=\frac{\lambda_w}{72W^4},\quad \gamma_v = \frac{\gamma_w+9W(3u_w+\lambda_w)}{27W^3} \\
    \gamma_{nq}&=\frac{4\gamma+3W(-\lambda_1+2\lambda_2)}{144W^3}\quad \gamma'_{nq}=\frac{-\lambda_1+2\lambda_2}{144W^4},\quad \gamma_{mnq}=\frac{-4\gamma+3W(\lambda_1+4\lambda_2)}{72W^3}\quad \gamma'_{mnq}=\frac{\lambda_1+4\lambda_2}{72W^4}.
\end{align}
When $W\gamma<0$, we showed in the previous section  that ferromagnetism is the leading instability, i.e., $T_\mathcal{M}>T_\mathcal{N}$. Symmetry prevents $\mathcal{M}$ and $\mathcal{M}^2$ to couple to a linear function of $\boldsymbol{\mathcal{N}}$ or $\boldsymbol{\mathcal{Q}}$. Thus, condensation of $\mathcal{M}$ does not accompany a secondary weak multipole, which is consistent with the stabilization of rotation-preserving FM (phase $1$) for $W\gamma<0$ in the phase diagram of the main text (Fig. 2(a)).

%On the other hand, the case of $W \gamma<0$ stabilizes either phase $3$ or phase $4$, owing to a competition between nematic (favoring phase $3$) and ferromagnetic (favoring phase $4$) degrees of freedom. This competition manifests in the anharmonic angle-dependent terms in $\mathcal{F}$. Quadratic terms cannot distinguish the two phases energetically, since terms which select between the two orientations of $\boldsymbol{\mathcal{N}}$ only appear at sixth-order, i.e., the sign of $\zeta$ in Eq. (\ref{simplified_FE}). 
On the other hand, the case of $W \gamma>0$ stabilizes either phase $3$ or phase $4$, depending on the sixth-order term in $\boldsymbol{\mathcal{N}}$. We minimize the free energy with respect to the ferromagnetic ($\mathcal{M}$), nematic ($\boldsymbol{\mathcal{Q}}$), and $\mathcal{V}$ order parameters. To obtain this term, it is convenient to first analyze the leading order behaviors of the secondary multipoles $\mathcal{M}$, $\boldsymbol{\mathcal{Q}}$, and $\mathcal{V}$ at an altermagnetic instability. Note that the TR-even composites couple to terms quadratic in $\boldsymbol{\mathcal{N}}$, whereas $\mathcal{M}$ couples to a term cubic in $\boldsymbol{\mathcal{N}}$. It follows that
\begin{equation}
    \mathcal{Q},\text{ }\mathcal{V}\propto\mathcal{N}^2 + \mathcal{O}(\mathcal{N}^4)\text{  and }\mathcal{M}\propto\mathcal{N}^3+\mathcal{O}(\mathcal{N}^5)
\end{equation}
where the coefficients (which may be zero) are set by $\theta$ and $\phi$. We can thus organize terms in $\mathcal{F}$ according to their scaling with $\mathcal{N}$. Indeed, the first angle-dependent term, proportional to $\gamma_{nq}\mathcal{N}^2\mathcal{Q}\cos(2\theta-\phi)$, appears at $\mathcal{O}(\mathcal{N}^4)$, and it is the only one that appears at that order. Hence, the relative orientation of the AM and nematic order parameters is set by the sign of $\gamma_{nq}$, where $\gamma_{nq}<0$ ($\gamma_{nq}>0$) gives $\phi=-2\theta$ ($\phi=-2\theta+\pi$). We thus eliminate $\phi$ from $\mathcal{F}$ and expand the secondary multipoles in a perturbative series in $\mathcal{N}$. It is necessary to compute the $\mathcal{O}(\mathcal{N}^4)$ corrections to $\mathcal{Q}$ and $\mathcal{V}$, as terms which are quadratic in $\mathcal{Q}$ or $\mathcal{V}$ can generate $\mathcal{N}^6$ terms. Upon solving $\partial\mathcal{F}/\partial \mathcal{Q}=\partial\mathcal{F}/\partial \mathcal{V}=\partial\mathcal{F}/\partial \mathcal{M}=0$ order-by-order in $\mathcal{N}$, we obtain

\begin{align}
    \mathcal{M}&=\frac{4\gamma_{mnq}\gamma_{nq}-9a_q\lambda_{mn}}{36a_m a_q}\mathcal{N}^3\sin3\theta + ...\\
    \begin{split}
    \mathcal{Q}&=\frac{|\gamma_{nq}|}{3a_q}\mathcal{N}^2+\frac{-36a_ma_n'a_qa_q'\gamma_{nq}+36a_ma_qa_vu_{nq}\gamma_{nq}-4a_qa_v\gamma_{mnq}^2\gamma_{nq}+36a_ma_n'a_q^2\gamma_{nq}'+9a_q^2a_v\gamma_{mnq}\lambda_{mn}}{216a_ma_q^3a_v}\mathcal{N}^4\\
    &+\frac{12a_m\gamma_{nq}(3a_3u_{nq}'-2\gamma_{nq}\gamma_q)+a_q\gamma_{mnq}(4\gamma_{mnq}\gamma_{nq}-9a_q\lambda_{mn})}{216a_ma_q^3}\mathcal{N}^4\cos 6\theta+...
    \end{split}\\ \mathcal{V}&= -\frac{a_n'}{2a_v}\mathcal{N}^2 + \frac{18a_n'a_n''a_q^2a_v-2a_v^2\gamma_{nq}(a_q'\gamma_{nq}-2a_q\gamma_{nq}')-9a_n'^2a_q^2\gamma_v}{36 a_q^2 a_v^3}\mathcal{N}^4+... 
\end{align}
Upon plugging these expressions into Eq. (\ref{simplified_FE_2}), we obtain the effective free energy
\begin{equation}
    \mathcal{F}(\boldsymbol{\mathcal{N}}) = \frac{a_n}{2}\mathcal{N}^2 + \frac{1}{4}\Big( u_n - \frac{a_n'^2}{2a_v}-\frac{2\gamma_{nq}^2}{9a_q} \Big) \mathcal{N}^4 + \frac{\zeta}{6}\mathcal{N}^6 \cos 6\theta + \frac{\zeta'}{6}\mathcal{N}^6 + ...
\end{equation}
Note that, as expected, the quartic and sixth-order terms are renormalized, and the latter are given by
\begin{align}
    \zeta &= \frac{16a_m\gamma_{nq}^2(9a_q u'_{nq} - 4\gamma_{nq}\gamma_q)+a_q(4\gamma_{mnq}\gamma_{nq}-9a_q \lambda_{mn})^2}{864a_ma_q^3} \\ \label{eq:zeta}
    \zeta' &= \frac{72a_m\big[9a_n'^2a_n''a_q^2a_v+2a_v^3u_{nq}\gamma_{nq}^2+2a_n'a_v^2\gamma_{nq}(a_q'\gamma_{nq}-2a_q\gamma_{nq}')+3a_n'^3a_q^2\gamma_v\big]+a_v^3(4\gamma_{mnq}\gamma_{nq}-9a_q\lambda_{mn})^2}{864a_ma_q^2a_v^4}.
\end{align}
 Recall that the sign of the angle-dependent term $\zeta$ favors either phase $3$ ($\zeta<0$) or phase $4$ ($\zeta>0$). The simplified regime of frozen out $\textbf{W}$ corresponds to taking $\gamma_{nq}=\gamma_q=\gamma_{mnq}=0$, which results in $\zeta=\frac{3\lambda^2}{32a_1}$, as in Eq. (\ref{N_only_FE_simplified}), with $\lambda_{mn}$ replaced by $\lambda$ and $a_m$ by $a_1$. 

We now recast everything in terms of the original parameters of the CDW-LC free energy of Eq. \ref{CDW_free_energy}, and assume $\gamma_w>0$ such that $W<0$. Analogous results hold for $\gamma_w<0$ and $W>0$. For clarity, we consider the simplified regime of $\lambda_1=\lambda_2=\lambda_3=0$, meaning that the LC and CDW order parameters only couple via the linear-quadratic coupling $\gamma$ in Eq. \ref{CDW_free_energy}. This is similar to the parameter range studied in the main text, where the quartic mixed terms are substantially smaller than all other Landau parameters. When $\gamma=0$ the CDW and LC degrees of freedom are independent, and in the LC-ordered state ($T<T_{\Phi,0}$), $\lambda_\Phi>0$ favors a $C_{3z}$-broken phase with $\textbf{W}=(W,W,W)$ and $\boldsymbol{\Phi}=(0,\Phi',0)$, respectively. Note that the suppression of the quartic mixed terms ($\lambda_1=\lambda_2=\lambda_3=0$) effectively enhances the symmetry of the CDW-LC configuration, and restoring them while keeping $\gamma=0$ stabilizes phase $4$: $(W,W,W)\to(W,W',W)$ and $(0,\Phi',0)\to(\Phi,\Phi',\Phi)$. Thus, we conclude that, in this regime, $\gamma=0$ favors phase 4. This can also be seen directly from Eq. (\ref{eq:zeta}): setting  $\lambda_1=\lambda_2=\lambda_3=0$ and $\gamma=0$ gives $\gamma_{nq}=0$ while $\lambda_{mn}\propto \lambda_\Phi$, such that $\zeta \propto \lambda_\Phi^2>0$.

When $\gamma<0$ is nonzero, the CDW and LC degrees of freedom are mixed, and as long as $|\gamma|$ is small, phase $4$ remains the global minimum as the sixth-order term $\zeta$ is positive. Once $\gamma$ reaches a critical value $\gamma^*<0$ (where $\zeta=0$), ferromagnetism is suppressed and phase $3$ (pure AM) is stabilized ($\zeta<0$). Solving $\zeta=0$ perturbatively in $\lambda_\Phi>0$ yields an expression on $\gamma^*$:

 \begin{equation}
 \gamma^*=-\Big(\frac{3}{2}\Big)^{2/3}\frac{3a_w+W(9Wu_w-\gamma_w)}{\sqrt[3]{a_\Phi(2\gamma_w-9W\lambda_w)}}\lambda_\Phi^{2/3}.
 \end{equation}
However, in this limit $a_\Phi=T_\mathcal{N}-T_{\Phi,0}=W\gamma^*/3>0$, which yields
\begin{equation}
    \gamma^*=-\Big(\frac{27}{2} \Big)^{1/4}\frac{[3a_w+W(9Wu_w-\gamma_w)]^{3/4}}{|W|^{1/4}(2\gamma_w-9W\lambda_w)^{1/4}}\lambda_\Phi^{1/2}
\end{equation}
 where $W<0$ is evaluated at $T_\mathcal{N}(\gamma^*)$. Note that, by definition, $\mathrm{sgn}(W) \mathrm{sgn}(\gamma_w) < 0$.

\subsection{Symmetry-enforced spin-splitting}

The non-zero TR-odd composites discussed in the previous sections will generate splitting between bands of opposite spins when SOC is present. Here, we derive the form of the $\textbf{k}$-dependent spin-splitting $\Delta E(\textbf{k})=E_\uparrow(\textbf{k})-E_\downarrow(\textbf{k})$ for the phases $1$, $3$, and $4$. Note that the latter two phases realize two nonzero TR-odd multipoles with different symmetries. Consequently, \textit{accidental} spin degeneracies can arise at low-symmetry $\textbf{k}$ due to a superposition of different irreps. The precise locations of the degenerate points in the Brillouin zone depend sensitively on the SOC scale, band index, the relative size of the two order parameters, and other microscopic quantities. Restricting to a phenomenological analysis, however, an expansion of the spin splitting near the $\Gamma$-point can be performed to leading harmonics by exploiting the fact that $\textbf{k}=k(\cos\theta_k,\sin\theta_k)$,  transforms as the  $m\Gamma_6^- = E_{1u}^{-}$ irrep and the out-of-plane spin polarization $\sigma_z$ transforms as the $m\Gamma_2^+=A_{2g}^-$ irrep. Then, by constructing polynomials of $\textbf{k}$ and $\sigma_z$ that transform as the irreps of the TR-odd composites that are non-zero in each of the three states, we find: 
\begin{align}
    \Delta E^{\text{phase 1}}(\textbf{k})&=\mathcal{M}(c_1+c_2|\textbf{k}|^6\cos 6\theta_k+...)\label{GS1_splitting}\\
    \Delta E^{\text{phase 3}}(\textbf{k})&=c_3\mathcal{N}|\textbf{k}|^2\sin(2\theta_k+n\pi/3)+c_4\Upsilon|\textbf{k}|^6\sin6\theta_k+...\label{GS3_splitting}\\
    \Delta E^{\text{phase 4}}(\textbf{k})&=c_5\mathcal{N}|\textbf{k}|^2\sin \left[2\theta_k+(2n+1)\pi/6 \right]+c_6\mathcal{M}+...\label{GS4_splitting}
\end{align}
where $n=0,1,...,5$ refers to the direction of the composite $\boldsymbol{\mathcal{N}}$. We introduced parameters $c_{1,2,3,4,5,6}$ that depend on microscopic details, as well as the first ``nodal'' $A_{1g}$ harmonic $\cos 6\theta_k$ to account for the possible emergence of accidental nodes in the presence of $\mathcal{M}$. We now describe the spin-splitting nodal structure of each state. As concrete examples for the different nodal structures, we quote the results shown in  Fig. \ref{fig:BZ_splitting}, which were obtained from the specific microscopic model presented in Section II of this Supplementary Material. 

\begin{figure*}
\includegraphics[width=0.85\columnwidth]{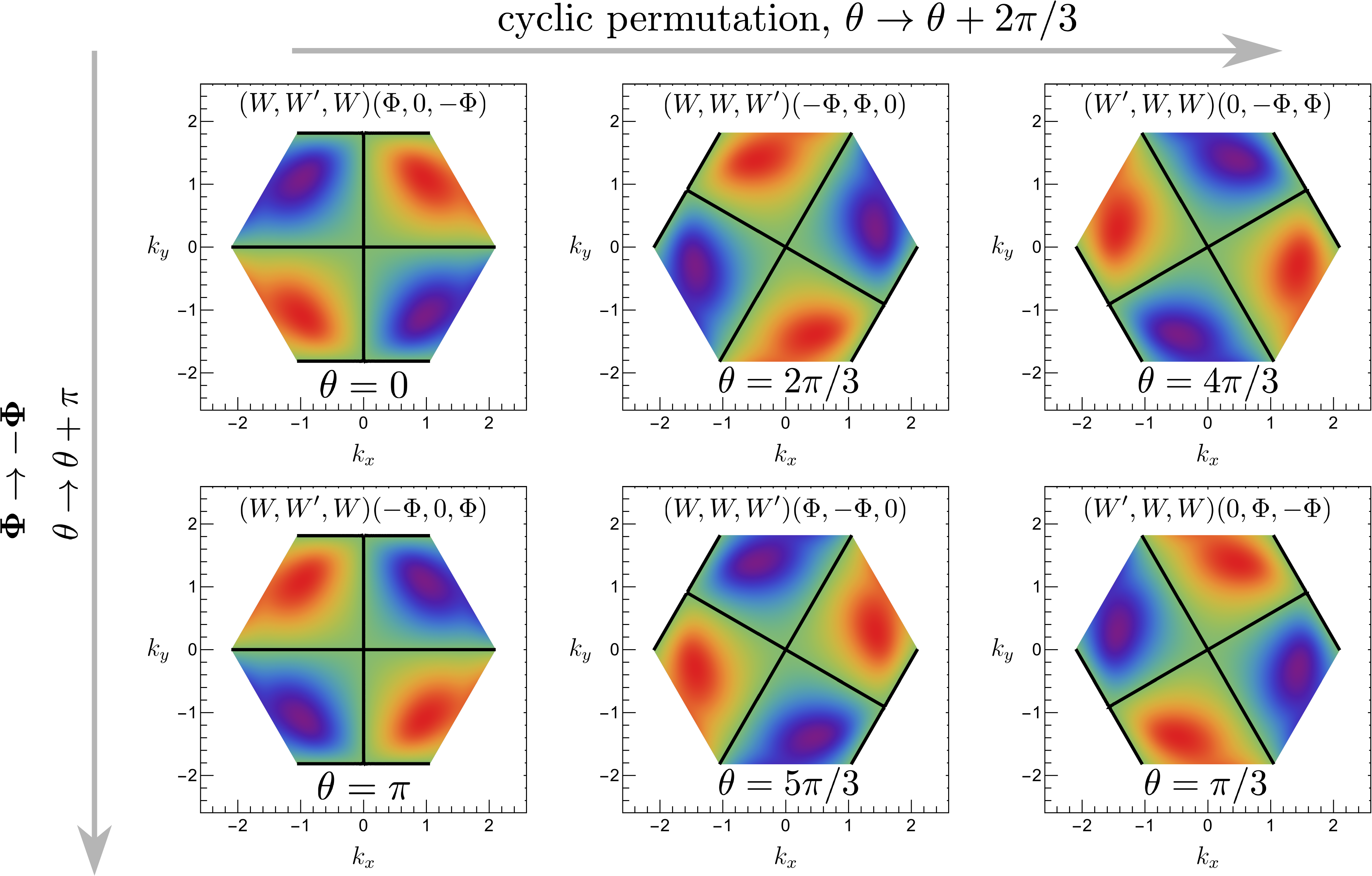}\caption{Momentum-dependence of the band structure spin splitting $\Delta E(\textbf{k})=E_\uparrow(\textbf{k})-E_\downarrow(\textbf{k})$ for each of the six domains associated with the pure AM phase (phase 3 in Table \ref{tab:GS_table}) . They are parameterized by the angle $\theta=n\pi/3$ with $n=0,1,...,5$, where $\boldsymbol{\mathcal{N}}=\mathcal{N}(\cos\theta,\sin\theta)$ is the $d$-wave AM order parameter, related to $\textbf{W}$ and $\boldsymbol{\Phi}$ via Eq. (\ref{N_def}). A cyclic permutation of the M-point results in a $2\pi/3$ $\textbf{k}$-space rotation with $\theta\to\theta+2\pi/3$ while TR flips the spins and takes $\theta\to\theta+\pi$. The split-bands shown here emerge from the $7$th band (counted by ascending energy: see Fig. \ref{fig:BZ_splitting} (a)) obtained by diagonalizing Eq.(\ref{big_bloch_hamiltonian}) for  $W=0.5$, $W'=0.5$, and $\Phi=0.5$. }\label{fig:BZ_splitting_AM}
\end{figure*}

\begin{itemize}
    \item \textbf{Phase $1$}: The FM state exhibits $D_{6h}$-preserving spin-splitting with no symmetry-enforced spin degeneracies. The ratio $c_2/c_1$ controls the relative strength of the non-nodal and nodal $s$-wave harmonics. When $c_1\gg c_2$, there are no accidental nodes since the non-nodal harmonic dominates; an example of this situation is shown in Fig. \ref{fig:BZ_splitting}(a). When $c_1$ and $c_2$ are comparable, accidental nodes can emerge away from the $\Gamma$-point for $|\textbf{k}|>|c_1/c_2|^{1/6}$; such a case is realized in Fig. \ref{fig:BZ_splitting}(b). 
    \item \textbf{Phase $3$}: The pure AM phase exhibits rotational symmetry breaking ($D_{6h}\to D_{2h}$) due to nematic order $\boldsymbol{\mathcal{Q}}\neq 0$, with a symmetry-enforced nodal plane at $\theta_k=m \pi/2-n\pi/6$ with $m=0,1,2,3$ that depends on the orientation of the leading $d$-wave harmonic $\boldsymbol{\mathcal{N}}$ encoded by $n$. As shown in the associated spin-split band structure of Fig. \ref{fig:BZ_splitting_AM}, in the presence of SOC each $n$ relates $\boldsymbol{\mathcal{N}}$ by three-fold rotations. Each corresponds to a particular cyclic permutation of the M-points that in turn rotates the band structure. However, because of the  subleading $i$-wave composite $\Upsilon$, additional accidental nodes can appear. Figs. \ref{fig:BZ_splitting}(c)-(d) illustrate the cases without accidental nodes and with accidental nodes, respectively.
    \item \textbf{Phase $4$}: The mixed AM $+$ FM phase exhibits nematic order ($D_{6h}\to D_{2h}$) and no symmetry-enforced nodes of the spin splitting. However, the subleading $s$-wave composite $\mathcal{M}$ can induce accidental nodes. Figs. \ref{fig:BZ_splitting}(e)-(f) show the cases without accidental nodes and with accidental nodes, respectively.
\end{itemize}

\subsection{Symmetries of the orbital magnetic phases: connection to magnetic and spin groups}
Altermagnetism has been defined using the spin group formalism, where spins transform as irreducible representations of SU$(2)$ rather than of the crystalline point group. Unlike spins, the orbital magnetic moments associated with loop currents do not have SU($2$) symmetry. Moreover,  these orbital moments are due to interatomic currents that are necessarily connected to the underlying crystalline group. As such, it is appropriate to classify these loop current states in terms of magnetic point groups (MPG). A classification of magnetic order parameters such as altermagnetism and ferromagnetism in terms of irreps of MPGs has been put forward by several works, see for instance Refs. \cite{Fernandes2024_AM,Scheurer2024,Ramires2026}. They can be understood as starting from a given spin group and then turning on a small SOC. We can therefore classify the composite $\textbf{Q}=0$ magnetic/non-magnetic states arising  from the 3Q CDW state by (single-valued) irreducible representations of the colorless MPG $6mmm.1'\cong D_{6h}\times\mathbb{Z}_2^T$, as shown in Table \ref{tab:table1}.\\

\begin{table}[h]
    \centering
    \begin{tabular}{|c|c|c|}
         \hline
         Phase &TRSB multipoles&TRS multipoles\\
         \hline
         \hline
         FM & $A_{2g}^-$ ($s$-wave), $\mathcal{O}(\Phi)$& - \\ \hline
         AFM & -& $E_{2g}^+$ (nematic), $\mathcal{O}(\Phi^2)$\\ \hline
         pure AM&$E_{2g}^-$ ($d$-wave), $\mathcal{O}(\Phi)$; $A_{1g}^-$ ($i$-wave), $\mathcal{O}(\Phi^3)$& $E_{2g}^+$ (nematic), $\mathcal{O}(\Phi^2)$\\ \hline
         mixed AM&$E_{2g}^-$ ($d$-wave), $\mathcal{O}(\Phi)$; $A_{2g}^-$ ($s$-wave), $\mathcal{O}(\Phi^3)$&$E_{2g}^+$ (nematic), $\mathcal{O}(\Phi^2)$\\ \hline
    \end{tabular}
    \caption{\small{List of leading and subleading multipoles, as indicated by the order in the LC order parameter $\boldsymbol{\Phi}$, that emerge in four different magnetic phases discussed in our manuscript. Here, TRSB and TRS refer to time-reversal symmetry-breaking and time-reversal symmetric orders. They are classified by irreps of the MPG  $6mmm.1'\cong D_{6h}\times\mathbb{Z}_2^T$  and identified using the nomenclature introduced in Ref. \cite{Fernandes2024_AM}.}}
    \label{tab:table1}
\end{table}

While orbital magnetic moments do not have  SU($2$) symmetry, in the presence of spin-orbit coupling (SOC), these orbital moments generate out-of-plane spin moments on the sites of the kagome lattice (see Fig. \ref{fig:unitcell_and_spinmagnetism}(e)-(g)). If we elevate them to fully SU($2$)-symmetric spins, we could use the spin group formalism to describe the ordered states. Because the spins do not have the same magnitude on the sites, a spin group description requires one to introduce additional ``colors'' corresponding to different magnitudes of the spin moment. For instance, in Fig. \ref{fig:unitcell_and_spinmagnetism}(e)-(f), besides spins up and down, one must introduce a spin of magnitude zero. In Fig. \ref{fig:unitcell_and_spinmagnetism}(g), one must introduce two types of spins up and down corresponding to the red and blue colored spins, which have different magnitudes. Thus, to accomplish such a description, it is necessary to generalize spin groups to accommodate multiple ``colors'', analogously to how Shubnikov (black-and-white) groups can be extended to polychromatic groups \cite{shubnikov_belov_colored_symmetry_1964} by including an additional operation of color permutation between crystallographic sites. Developing such a formalism, while interesting, is beyond the scope of our work.

\section{Details of the tight-binding model}
\begin{figure*}
\includegraphics[width=1.0\columnwidth]{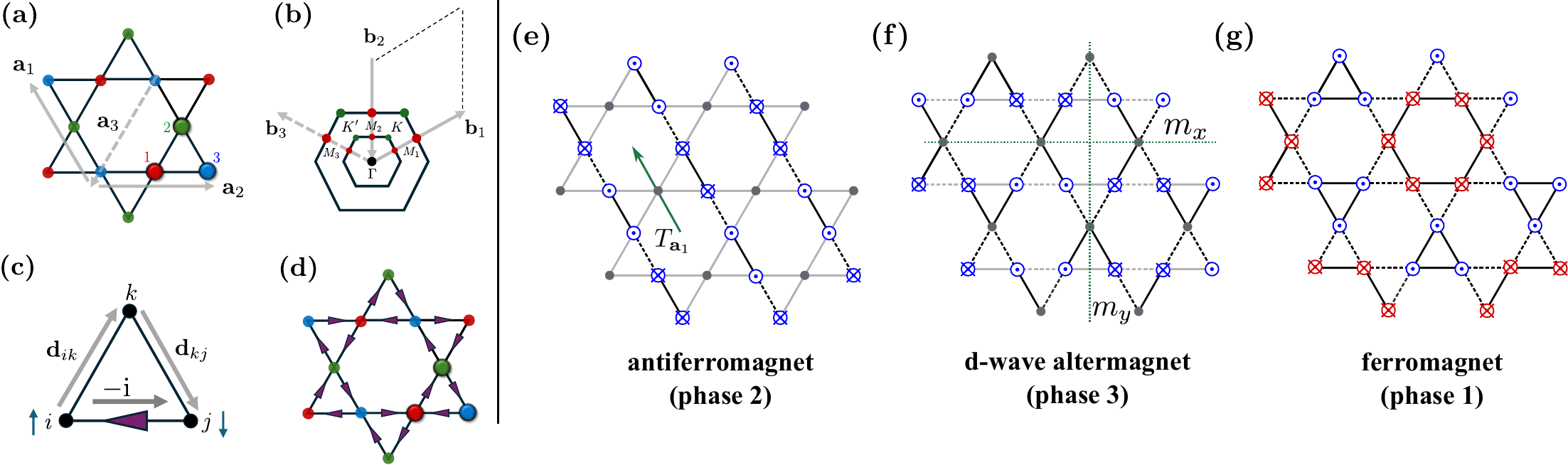}\caption{\textbf{(a)}: The kagome lattice with labeled sublattices $1$, $2$, and $3$ and the Bravais lattice vectors $\textbf{a}_{1,2,3}$. \textbf{(b)}: First Brillouin zone (BZ) of the kagome lattice with reciprocal vectors $\textbf{b}_{1,2,3}$ and the high-symmetry M-points and K-points of the original lattice (outer hexagon) and those of the charged-ordered lattice (inner hexagon). \textbf{(c)}, \textbf{(d)}: Schematic of the SOC term in the tight-binding model. \textbf{(e)-}\textbf{(g)}: Spin configuration induced by SOC in the presence of primary orbital magnetic order. The crossed and dotted circles of the same color correspond to in-plane and out-of-plane moments of equal magnitude.  \textbf{(e)}: Phase $2$ with AFM character, where opposite-spin magnetic sublattices are connected by translation. \textbf{(f)}: Phase $3$ with primary $d$-wave AM character, where opposite-spin magnetic sublattices are connected by mirrors $m_y$ and $m_x$. \textbf{(g)}: Phase $1$ with primary FM character; there is no compensation of spins.} \label{fig:unitcell_and_spinmagnetism}
\end{figure*}

In this section, we show the details of the construction of an effective nearest-neighbor tight-binding model on a kagome lattice with spin-orbit coupling (SOC) and coupled CDW-LC orders. We include terms that couple $W_i$ and $\Phi_i$ to fermionic bilinears in accordance with their transformation properties. For nonzero SOC, we numerically compute the spin-polarized band structure, and obtain Fig. 3 in the main text for the FM, $d$-wave AM, and AFM cases (corresponding to phases 1, 3, and 2 in Table \ref{tab:GS_table}, respectively).

To begin, we define a set of three overcomplete Bravais lattice vectors $\textbf{a}_{1,2,3}$ that span the plane, along with a set of overcomplete reciprocal lattice vectors $\textbf{b}_{1,2,3}$, as shown in Fig. \ref{fig:unitcell_and_spinmagnetism} (a)-(b). Setting the lattice constant to unity, they are given in Cartesian coordinates by 
\begin{align}
    \textbf{a}_{1,3}&=\frac{1}{2}(-1,\pm\sqrt{3}),\quad \textbf{a}_2=(1,0) \\
    \textbf{b}_{1,3}&=\frac{2\pi}{\sqrt{3}}(\pm\sqrt{3},1),\quad \textbf{b}_2=-\frac{4\pi}{\sqrt{3}}(0,1).
\end{align}
This definition yields 
\begin{equation}
    \textbf{b}_i\cdot \textbf{a}_j=2\pi\sum_{n=1}^3\epsilon_{ijn}
\end{equation}
where $\epsilon_{ijn}$ is the Levi-Civita symbol. Crucially, this inner product is nonzero if and only if $i\neq j$, and it satisfies the anti-symmetry property $\textbf{b}_i\cdot\textbf{a}_j=-\textbf{b}_j\cdot \textbf{a}_i$ along with $\textbf{b}_1\cdot\textbf{a}_2=\textbf{b}_2\cdot \textbf{a}_3=\textbf{b}_3\cdot \textbf{a}_1=2\pi$. It is convenient to introduce ($i,j,l$) as a cyclic permutation of ($1,2,3$) which is determined by $i\in\{1,2,3\}$. With this constraint, the latter condition can be written as $\textbf{b}_i\cdot \textbf{a}_j=1$. Unless otherwise stated, all following instances of the ordered triplet ($i,j,l$) satisfy this constraint. Since $\textbf{a}_1+\textbf{a}_2+\textbf{a}_3=0$ and $\textbf{b}_1+\textbf{b}_2+\textbf{b}_3=0$, if $\textbf{a}_i$ corresponds to a vertex of an equilateral triangle, then $\textbf{b}_i$ corresponds to the opposite leg of the triangle. Consequently, the following relationship holds between direct and reciprocal vectors:
\begin{align}
    \textbf{a}_i=-\frac{\textbf{b}_j-\textbf{b}_l}{4\pi},\quad \textbf{b}_i=\frac{4\pi}{3}(\textbf{a}_j-\textbf{a}_l)
\end{align}

\subsection{Real space}

Letting $r$, $r'$ denote vertices (not vectors) of the kagome lattice, we define fermions $\psi_{r\alpha}^{\phantom{}}$, $\psi_{r\alpha}^\dagger$ in the site basis. In this basis, the nearest-neighbor tight-binding Hamiltonian describing a kagome lattice with SOC is given by 
\begin{equation}
\mathcal{H}_0=-t\sum_{\langle rr' \rangle}\psi_{r\alpha}^\dagger \psi_{r'\alpha}^{\phantom{}} + \mathrm{i}\lambda_\text{soc}\sum_{\langle rr' \rangle}\nu_{rr'}\sigma_z^{\alpha\beta}\psi_{r\alpha}^\dagger\psi_{r'\beta}^{\phantom{}}+\text{H.c.}
\end{equation}
where repeated spin indices $\alpha$, $\beta\in{\{\uparrow,\downarrow\}}$ are implicitly summed, and $t$ and $\lambda_\text{soc}$ are the hopping and the SOC energy scale, respectively. Additionally, $\nu_{rr'}=\pm 1$ is determined by the handedness of the path starting at $r$ and ending at $r'$ along its unique shared neighbor $r''$. It may alternatively be represented as $\nu_{rr'}=\frac{8}{\sqrt{3}}(\textbf{d}_{rr''}\times\textbf{d}_{r''r})\cdot \hat{z}$ where $\textbf{d}_{rr''}$ ($\textbf{d}_{r''r'}$) is a vector with tail at $r$ ($r''$) and tip at $r''$ ($r'$), see Fig. \ref{fig:unitcell_and_spinmagnetism}(c)-(d). The extra factor of $8/\sqrt{3}$ comes from the fact that in-plane $\textbf{d}$'s are each \textit{half} the length of a Bravais unit cell, such that $|\textbf{d}_{rr''}\times\textbf{d}_{r''r}|=(\frac{1}{2})^2\sin\frac{\pi}{3}=\frac{\sqrt{3}}{8}$. At first sight, it may seem that the notation $\langle rr'\rangle$ is ambiguous, as it merely refers to an unordered nearest-neighbor pair. However, for a given pair $\langle rr' \rangle$ the operation $r \leftrightarrow r'$ takes $\nu_{rr'}\to \nu_{r'r}=-\nu_{rr'}$ and $(\psi_{r\uparrow}^\dagger\psi_{r'\uparrow}^{\phantom{}}-\psi_{r\downarrow}^\dagger\psi_{r'\downarrow}^{\phantom{}})\to -(\psi_{r\uparrow}^\dagger\psi_{r'\uparrow}^{\phantom{}}-\psi_{r\downarrow}^\dagger\psi_{r'\downarrow}^{\phantom{}})$. Hence, it takes $\mathrm{i}\lambda\sum_{\langle rr' \rangle}\nu_{rr'}(\psi_{r\uparrow}^\dagger\psi_{r'\uparrow}^{\phantom{}}-\psi_{r\downarrow}^\dagger\psi_{r'\downarrow}^{\phantom{}})$ to itself, and the order in which one chooses $r$ and $r'$ does not matter. Importantly, $\sigma_z$ still remains a good quantum number even in the presence of SOC. 

Before proceeding, we note that, in the presence of SOC, the orbital magnetic moments shown in Fig. 2(b)-(d) of the main text will be accompanied by spin order in the sites of the kagome lattice. The spin-order pattern can be directly obtained from group theory, using the MSG symmetries of each state. They are illustrated in Figs. \ref{fig:unitcell_and_spinmagnetism}(e)-(g) for the case of the AFM (phase 2), $d$-wave AM (phase 3) and FM (phase 1) states, respectively.

We now distinguish physically independent order parameters transforming as the same irrep. Up to nearest-neighbor additions to $\mathcal{H}_0$, there are two independent CDW order parameters transforming like $M_1^+$: a \textit{site} order parameter $\textbf{w}=(w_1,w_2,w_3)$ representing local modulations of the onsite energy, and a \textit{bond} order parameter $\textbf{W}=(W_1,W_2,W_3)$, representing nearest-neighbor modulations of the hopping amplitude $t$. Because they are necessarily proportional to each other, we can write $\mathbf{w}=\eta \mathbf{W}$, with some constant $\eta$. To capture their effects on the electronic spectrum, we introduce $\mathcal{H}_\text{CDW}$ and $\mathcal{H}_\text{CDW}'$. There is also one LC order parameter transforming like $mM_2^+$, $\boldsymbol{\Phi}=(\Phi_1,\Phi_2,\Phi_3)$, representing nearest-neighbor modulations of the phase of the hopping parameter, which we capture by $\mathcal{H}_\text{LC}$. The three components of each order parameter correspond to modulations perpendicular to $\textbf{a}_1$, $\textbf{a}_2$, and $\textbf{a}_3$ axes, thus associated with wave-vectors $\textbf{Q}_1$, $\textbf{Q}_2$, and $\textbf{Q}_3$, respectively (recall that $\textbf{Q}_i=\textbf{b}_i/2$). The full Hamiltonian with bond CDW, site CDW, and LC contributions thus become:

\begin{equation}
\begin{split}
\mathcal{H}_\text{tot}&=\mathcal{H}_0+\mathcal{H}_\text{LC}+\mathcal{H}_\text{CDW}+\mathcal{H}_\text{CDW}'\\
\mathcal{H}_\text{CDW}&=g_W\sum_{\langle rr'\rangle}\textbf{W}\cdot \textbf{t}_{rr'}\psi_{r\alpha}^\dagger \psi_{r'\alpha}^{\phantom{}}+\text{H.c.}\\
\mathcal{H}_\text{CDW}'&=g_W'\sum_{r}\textbf{w}\cdot \boldsymbol{\rho}_r \psi_{r\alpha}^\dagger \psi_{r\alpha}^{\phantom{}}+\text{H.c.}\\
\mathcal{H}_\text{LC}&=\mathrm{i}g_\Phi \sum_{\langle rr' \rangle}\boldsymbol{\Phi}\cdot \boldsymbol{\lambda}_{rr'}\psi_{r\alpha}^\dagger \psi_{r'\alpha}^{\phantom{}}+\text{H.c.}
\end{split}
\end{equation}

Here, $g_W$ and $g_\Phi$ are energy scales that describe the coupling strength of $\textbf{W}$ and $\boldsymbol{\Phi}$ to the hopping amplitude and phase, respectively, and $g_W'$ describes the coupling strength of $\textbf{w}$ to the onsite energy. For our purposes, we absorb these coupling constants into the definition of the corresponding order parameters and set $g_W=g_W'=g_\Phi=1$. Additionally, $\textbf{t}_{rr'}$ and $\boldsymbol{\lambda}_{rr'}$ are three-component ``vectors'' of real-symmetric and real-antisymmetric matrices, respectively, with the property that $\textbf{t}_{rr'}$, $\boldsymbol{\lambda}_{rr'}\neq 0$ if and only if $r$ and $r'$ are nearest neighbors. There is also an on-site real three-component $\boldsymbol{\rho}_i$ with $\textbf{w}\cdot\boldsymbol{\rho}_{r}=\delta\epsilon_r$ describing the onsite energy modulation. The objects $\boldsymbol{\rho}_r$, $\textbf{t}_{rr'}$, and $\boldsymbol{\lambda}_{rr'}$ are constrained by symmetry, such that the corresponding fermionic bilinears transform like $M_1^+$, $M_1^+$, and $mM_2^+$, respectively. We will give their expressions in a more convenient basis below.

It is more instructive to write $\mathcal{H}$ in the sublattice basis, described by fermions $d_{i\textbf{r},\alpha}$, where $i\in \{1,2,3 \}$ is the sublattice index and $\textbf{r}=n_1\textbf{a}_1+n_2\textbf{a}_2$ ($n_{1,2}\in\mathbb{Z}$) is a Bravais lattice vector. For concreteness, we let $d$ be trivially transforming ($A_g$) atomic-orbitals of the site-symmetry group $D_{2h}$, which encompass not only $s$-orbitals, but also higher angular momentum orbitals such as $d_{z^2}$ and $d_{x^2-y^2}$. We choose the three sublattices such that they are in one-to-one correspondence with the direct lattice vectors $\textbf{a}_{1,2,3}$, as illustrated in Fig. \ref{fig:unitcell_and_spinmagnetism}(a). In particular, recalling that ($i,j,l$) is a cyclic permutation of $(1,2,3)$, they are defined such that a sublattice $j$ state at site $\textbf{R}_j$ and a sublattice $l$ state at site $\textbf{R}_l$ within a single unit-cell are connected by $\textbf{a}_i/2$, i.e., $\textbf{R}_j-\textbf{R}_l=\textbf{a}_i/2$. With this machinery in mind, we express the CDW and LC order parameters, transforming as $M_1^+$ and $mM_2^+$, respectively, as a bilinears in the creation and annihilation operators of the atomic-orbitals $d$. To this end, we demonstrate how this works for the LC order parameter. Using Eq. (2) of the main text, we have

\begin{equation}
    \Phi_i=\mathrm{i}\sum_{\textbf{r},\alpha}\langle e^{\mathrm{i}\textbf{Q}_i\cdot\textbf{r}}d_{j\textbf{r},\alpha}^\dagger(d_{l\textbf{r},\alpha}^{\phantom{}}-d_{l\textbf{r}+\textbf{a}_i,\alpha}^{\phantom{}}) -\text{H.c.}\rangle 
    \label{phi}
\end{equation}
Upon orienting the crystallographic axes in accordance with Fig. \ref{fig:unitcell_and_spinmagnetism}(a), we need to determine the transformation properties of $\Phi_i$ for the two rotations $2_{001}$, $2_{010}$, and inversion $\bar{1}$ that make up the little co-group of the M-point, along with time-reversal $\mathcal{T}$. For the spatial operations, we leave off the spin index $\alpha$ for simplicity and find

\begin{align*}
2_{001}\text{, }\bar{1}
&: d_{1,\mathbf{r}} \mapsto d_{1,\mathbf{r}+\mathbf{a}_1}
&\qquad
2_{010}
&: d_{1,\mathbf{r}} \mapsto d_{1,\mathbf{r}+\mathbf{a}_1} \\
&: d_{2,\mathbf{r}} \mapsto d_{2,\mathbf{r}-\mathbf{a}_2}
&
&: d_{2,\mathbf{r}} \mapsto d_{2,\mathbf{r}-\mathbf{a}_2} \\
&: d_{3,\mathbf{r}} \mapsto d_{3,\mathbf{r}-\mathbf{a}_2+\mathbf{a}_1}
&
&: d_{3,\mathbf{r}} \mapsto d_{3,\mathbf{r}-\mathbf{a}_2+\mathbf{a}_1}
\end{align*}

Using Eq. (\ref{phi}), we thus find that  $2_{001}[\boldsymbol{\Phi}]=\bar{1}[\boldsymbol{\Phi}]=\boldsymbol{\Phi}$ whereas $2_{010}$ acts non-trivially on $\boldsymbol{\Phi}$:

\begin{equation}
2_{010}[\boldsymbol{\Phi}] = -\begin{pmatrix}
        1 & 0 & 0\\
        0 & 0 & 1\\
        0 & 1 & 0
    \end{pmatrix}\begin{pmatrix}
        \Phi_1\\\Phi_2\\\Phi_3
    \end{pmatrix} = \begin{pmatrix}
        -\Phi_1\\-\Phi_3\\-\Phi_2
    \end{pmatrix}
\end{equation}

Using the space group representation tables  of $P6/mmm$, available for instance in the Bilbao Crystallographic Server, we conclude that the above transformation properties correspond to the $M_2^+$ irrep of $P6/mmm$. On the other hand, $\mathcal{T}$ merely flips the spin of each electron $d_\uparrow \to d_\downarrow$ and complex conjugates Eq. (\ref{phi}). Therefore $\mathcal{T}[\boldsymbol{\Phi}]=-\boldsymbol{\Phi}$, thereby confirming that $\boldsymbol{\Phi}$ transforms as $mM_2^+$. It is straightforward to repeat this analysis for $M_1^+$-transforming $\textbf{W}$ using Eq. (1) of the main text.

This procedure can also be repeated for electronic states that transform as nontrivial irreps of the site-symmetry group $D_{2h}$. Such an analysis was in fact done by two of us in Ref. \cite{christensen2022toroidal}. The outcome is that, provided the electron operators involve the same atomic-orbital, the order parameter in Eq. (\ref{phi}) always transforms as $mM_2^+$.

Thus, using these atomic-orbitals, the full Hamiltonian acquires a particularly simple form in real space:

\begin{equation}
    \mathcal{H}=\sum_{i,\textbf{r},\alpha}\big[(-t+\mathrm{i}\lambda_\text{soc}\sigma_z^{\alpha\alpha})d_{j\textbf{r},\alpha}^\dagger(d_{l\textbf{r},\alpha}^{\phantom{}}+d_{l\textbf{r}+\textbf{a}_i,\alpha}^{\phantom{}})+(W_i+\mathrm{i}\Phi_i)e^{\mathrm{i}\textbf{Q}_i\cdot \textbf{r}}d_{j\textbf{r},\alpha}^\dagger(d_{l\textbf{r},\alpha}^{\phantom{}}-d_{l\textbf{r}+\textbf{a}_i,\alpha})+w_ie^{\mathrm{i}\textbf{Q}_i\cdot\textbf{r}}d_{i\textbf{r},\alpha}^\dagger d_{i\textbf{r},\alpha}^{\phantom{}} \big]+\text{H.c.}
\end{equation}
Since $\mathcal{H}$ is diagonal in spin, as $\sigma_z$ remains a good quantum number even in the presence of SOC, it is convenient to define a complex variable $\tau \equiv t - \mathrm{i}\lambda_\text{soc}$ such that conjugating $\tau \to\bar{\tau}=t+\mathrm{i}\lambda_\text{soc}$ amounts to effectively flipping the $z$-component of the spin. We also define $Z_i\equiv W_i+\mathrm{i}\Phi_i$, as the two order parameters $W_i$ and $\Phi_i$ always appear in tandem. It is thus convenient to define  the effective spinless Hamiltonian
\begin{equation}
\tilde{\mathcal{H}}\left[ \tau \right]=\sum_{i,\textbf{r}}\big[-\tau d_{j\textbf{r}}^\dagger(d_{l\textbf{r}}^{\phantom{}}+d_{l\textbf{r}+\textbf{a}_i}^{\phantom{}})+Z_ie^{\mathrm{i}\textbf{Q}_i\cdot \textbf{r}}d_{j\textbf{r}}^\dagger(d_{l\textbf{r}}^{\phantom{}}-d_{l\textbf{r}+\textbf{a}_i})+w_ie^{\mathrm{i}\textbf{Q}_i\cdot\textbf{r}}d_{i\textbf{r}}^\dagger d_{i\textbf{r}}^{\phantom{}} \big]+\text{H.c.}
\label{real_space_hamiltonian}
\end{equation}
The spin-up eigenvalues/eigenstates of $\mathcal{H}$ thus correspond to the eigenvalues/eigenstates of $\tilde{\mathcal{H}}\left[ \tau \right]$, whereas the spin-down ones are obtained from  $\tilde{\mathcal{H}}\left[\bar{\tau} \right]$. In the following, we thus focus on the properties of $\tilde{\mathcal{H}}$.

\subsection{Momentum space}
In this subsection, we analyze  $\tilde{\mathcal{H}}$ in momentum space. Note that CDW-LC order quadruples the unit-cell and thus reduces the Brillouin zone in one quarter. As such, it is useful to define the \textit{original} first Brillouin zone (labeled BZ) and the dimerized \textit{mini} first Brillouin zone (labeled BZ') given by the smaller hexagon in Fig. \ref{fig:unitcell_and_spinmagnetism}(b). They are given in terms of the reciprocal vectors as
\begin{align}
    \text{BZ}&:=\{p_1\textbf{b}_1+p_2 \textbf{b}_2|p_{1,2}\in[0,1] \}\\
    \text{BZ'}&:=\{p_1\textbf{b}_1+p_2 \textbf{b}_2|p_{1,2}\in[0,1/2] \}
\end{align}
along with \textit{new} high-symmetry points $\bar{M}_i$, $\bar{K}$, and $\bar{K}'$, given by $\textbf{Q}_{\bar{M}_i}\equiv\textbf{Q}_{M_i}/2$, $\textbf{Q}_{\bar{K}}\equiv\textbf{Q}_K/2$, and $\textbf{Q}_{\bar{K}'}:\equiv \textbf{Q}_{K'}/2$. Upon taking a Fourier transform of the sublattice fermion operators 
\begin{equation}
    d_{i\textbf{r}}=\frac{1}{\sqrt{N}}\sum_{\textbf{k}\in\text{BZ}}e^{\mathrm{i}\textbf{k}\cdot\textbf{r}}d_{i\textbf{k}}
\end{equation}
we write Eq. (\ref{real_space_hamiltonian}) in momentum space:

\begin{equation}
 \tilde{\mathcal{H}}=\sum_{\textbf{k}\in\text{BZ, }i}\big[ -\tau(1+e^{\mathrm{i}k_i})d_{j\textbf{k}}^\dagger d_{l\textbf{k}}^{\phantom{}}+Z_i(1-e^{\mathrm{i}k_i})d_{j\textbf{k}+\textbf{Q}_i}^\dagger d_{l\textbf{k}}^{\phantom{}}+w_id_{i\textbf{k}+\textbf{Q}_i}^\dagger d_{i\textbf{k}}^{\phantom{}}\big]+\text{H.c.}
\label{momentum_space_hamiltonian}
\end{equation}
Here, $k_i\equiv \textbf{k}\cdot \textbf{a}_i$ so that $k_1+k_2+k_3=0$. Obviously, the case of $Z_i$, $w_i\neq0$ induces off-diagonal-in-$\textbf{k}$ terms when the sum runs over the entire original Brillouin zone. The Bloch Hamiltonian as we express below is a $12$-by-$12$ $\textbf{k}$-dependent matrix, with the twelve bands coming from the dimerization of an originally $3$-sublattice model and with $\textbf{k}$ running over the restricted mini Brillouin zone. Before doing this, however, we make a convenient basis change. Specifically, we consider a transformation of the form 
\begin{equation}
    d_{i\textbf{k}}^{\phantom{}}=e^{-\mathrm{i}{\theta_i(\textbf{k})}}\tilde{d}_{i\textbf{k}}^{\phantom{}}
\end{equation}
that amounts to a similarity transformation (diagonal in the sublattice basis) of the Bloch Hamiltonian. Hence, the bilinears pick up phases:
\begin{align}
    &d_{j\textbf{k}}^\dagger d_{l\textbf{k}}^{\phantom{}}= e^{\mathrm{i}(\theta_j(\textbf{k})-\theta_l(\textbf{k}))} \tilde{d}_{j\textbf{k}}^\dagger \tilde{d}_{l\textbf{k}}^{\phantom{}}\\
    &d_{j\textbf{k}+\textbf{Q}_i}^\dagger d_{l\textbf{k}}^{\phantom{}}= e^{\mathrm{i}(\theta_j(\textbf{k}+\textbf{Q}_i)-\theta_l(\textbf{k}))} \tilde{d}_{j\textbf{k}+\textbf{Q}_i}^\dagger \tilde{d}_{l\textbf{k}}^{\phantom{}}\\
    &d_{i\textbf{k}+\textbf{Q}_i}^\dagger d_{i\textbf{k}}^{\phantom{}}= e^{\mathrm{i}(\theta_i(\textbf{k}+\textbf{Q}_i)-\theta_i(\textbf{k}))} \tilde{d}_{i\textbf{k}+\textbf{Q}_i}^\dagger \tilde{d}_{i\textbf{k}}^{\phantom{}}
\end{align}
Upon choosing 
\begin{equation}
    \theta_i(\textbf{k})=\Big(\textbf{R}+\frac{\textbf{b}_i}{8\pi}\Big)\cdot \textbf{k}
\end{equation}
where $\textbf{R}$ is some undetermined real-space vector, we can express Eq. (\ref{momentum_space_hamiltonian}) in terms of this new basis as
\begin{equation}
   \tilde{\mathcal{H}}=\sum_{\textbf{k}\in\text{BZ, }i}\big[ -2\tau\cos\frac{k_i}{2}\tilde{d}_{j\textbf{k}}^\dagger \tilde{d}_{l\textbf{k}}^{\phantom{}}+e^{\mathrm{i}\alpha_i}\big(-2Z_i\sin\frac{k_i}{2}\tilde{d}_{j\textbf{k}+\textbf{Q}_i}^\dagger \tilde{d}_{l\textbf{k}}^{\phantom{}}+w_i\tilde{d}_{i\textbf{k}+\textbf{Q}_i}^\dagger \tilde{d}_{i\textbf{k}}^{\phantom{}}\big)\big]+\text{H.c.}
\label{momentum_space_hamiltonian_2}
\end{equation}
along with an \textbf{R}-dependent phase
\begin{equation}
    \alpha_i \equiv \textbf{R}\cdot \textbf{Q}_i+\frac{\pi}{3}.
\end{equation}
This new basis affords one useful aspect: the complex $\textbf{k}$-dependent prefactors of the $\tilde{d}^\dagger \tilde{d}^{\phantom{}}$ bilinears appear as complex constants ($\tau$, $e^{\mathrm{i{\alpha_i}}}Z_i$, $e^{\mathrm{i{\alpha_i}}}w_i$) multiplied by (real) functions of $\textbf{k}$. Consequently, the entries of the $12$-by-$12$ Bloch Hamiltonian matrix, which are filled up by these functions, have complex phases that are necessarily independent of $\textbf{k}$. As we show below, this turns out to be crucial in showing the spin degeneracy of bands in the entire reduced Brillouin zone for the ``low-symmetry'' AFM state $\textbf{W}=(W,0,0)$, $\boldsymbol{\Phi}=(0,\Phi,\Phi')$, corresponding to phase $5$ with MSG $P2/m.1_a'$. This is actually a subleading instability compared to the higher symmetry AFM state (corresponding to $\Phi=\Phi'$) studied in the main text (phase $2$ with MSG $Cmmm.1_a'$). However, the latter is merely a special case of the former, and the same analysis follows.

For completeness, we give the Bloch Hamiltonian $\tilde{H}(\textbf{k})$ by introducing a $12$-component operator with $\textbf{k}\in\text{BZ'}$
\begin{equation}
    \textbf{d}_\textbf{k}^\dagger:=
    \begin{pmatrix}
    \tilde{d}_{1\textbf{k}}^\dagger 
    &\tilde{d}_{2\textbf{k}}^\dagger 
    &\tilde{d}_{3\textbf{k}}^\dagger &\tilde{d}_{1\textbf{k}+\textbf{Q}_1}^\dagger &\tilde{d}_{2\textbf{k}+\textbf{Q}_1}^\dagger &\tilde{d}_{3\textbf{k}+\textbf{Q}_1}^\dagger &\tilde{d}_{1\textbf{k}+\textbf{Q}_2}^\dagger &\tilde{d}_{2\textbf{k}+\textbf{Q}_2}^\dagger &\tilde{d}_{3\textbf{k}+\textbf{Q}_2}^\dagger &\tilde{d}_{1\textbf{k}+\textbf{Q}_3}^\dagger &\tilde{d}_{2\textbf{k}+\textbf{Q}_3}^\dagger &\tilde{d}_{3\textbf{k}+\textbf{Q}_3}^\dagger \end{pmatrix}
\end{equation}
such that Eq. (\ref{momentum_space_hamiltonian_2}) can be written for arbitrary $\textbf{R}$:
\begin{equation}
    \tilde{\mathcal{H}}=\sum_{\textbf{k}\in\text{BZ'}}\textbf{d}_\textbf{k}^\dagger \tilde{H}(\textbf{k})\textbf{d}_\textbf{k}
\end{equation}
For the choice of $\textbf{R}=-\frac{\textbf{b}_1}{8\pi}$, corresponding to $\alpha_1=\pi$ and $\alpha_{2,3}=\frac{\pi}{2}$, $\tilde{H(\textbf{k})}$ takes the matrix form
\begin{equation}
\scalemath{0.7}{\begin{pmatrix}
 0 & -2 \bar{\tau} \cos \frac{k_3}{2} & -2 \tau  \cos \frac{k_2}{2} & w_1 & 0 & 0 & 0 & 0 & 2 \mathrm{i} \bar{Z}_2 \sin \frac{k_2}{2} & 0 & 2 \mathrm{i} Z_3 \sin \frac{k_3}{2} & 0 \\
 -2 \tau  \cos \frac{k_3}{2} & 0 & -2 \bar{\tau} \cos \frac{k_1}{2} & 0 & 0 & 2 Z_1 \sin \frac{k_1}{2} & 0 & \mathrm{i} w_2 & 0 & 2 \mathrm{i} \bar{Z}_3 \sin \frac{k_3}{2} & 0 & 0 \\
 -2 \bar{\tau} \cos \frac{k_2}{2} & -2 \tau  \cos \frac{k_1}{2} & 0 & 0 & -2 \bar{Z}_1 \sin \frac{k_1}{2} & 0 & 2 \mathrm{i} Z_2 \sin \frac{k_2}{2} & 0 & 0 & 0 & 0 & \mathrm{i} w_3 \\
 w_1 & 0 & 0 & 0 & 2 \bar{\tau} \sin \frac{k_3}{2} & 2 \tau  \sin \frac{k_2}{2} & 0 & 2 \mathrm{i} Z_3 \cos \frac{k_3}{2} & 0 & 0 & 0 & 2 \mathrm{i} \bar{Z}_2 \cos \frac{k_2}{2} \\
 0 & 0 & -2 Z_1 \sin \frac{k_1}{2} & 2 \tau  \sin \frac{k_3}{2} & 0 & -2 \bar{\tau} \cos \frac{k_1}{2} & 2 \mathrm{i} \bar{Z}_3 \cos \frac{k_3}{2} & 0 & 0 & 0 & -\mathrm{i} w_2 & 0 \\
 0 & 2 \bar{Z}_1 \sin \frac{k_1}{2} & 0 & 2 \bar{\tau} \sin \frac{k_2}{2} & -2 \tau  \cos \frac{k_1}{2} & 0 & 0 & 0 & -\mathrm{i} w_3 & 2 \mathrm{i} Z_2 \cos \frac{k_2}{2} & 0 & 0 \\
 0 & 0 & -2 \mathrm{i} \bar{Z}_2 \sin \frac{k_2}{2} & 0 & -2 \mathrm{i} Z_3 \cos \frac{k_3}{2} & 0 & 0 & -2 \bar{\tau} \sin \frac{k_3}{2} & 2 \tau  \cos \frac{k_2}{2} & w_1 & 0 & 0 \\
 0 & -\mathrm{i} w_2 & 0 & -2 \mathrm{i} \bar{Z}_3 \cos \frac{k_3}{2} & 0 & 0 & -2 \tau  \sin \frac{k_3}{2} & 0 & 2 \bar{\tau} \sin \frac{k_1}{2} & 0 & 0 & 2 Z_1 \cos \frac{k_1}{2} \\
 -2 \mathrm{i} Z_2 \sin \frac{k_2}{2} & 0 & 0 & 0 & 0 & \mathrm{i} w_3 & 2 \text{$\bar{\tau}$} \cos \frac{k_2}{2} & 2 \tau  \sin \frac{k_1}{2} & 0 & 0 & 2 \bar{Z}_1 \cos \frac{k_1}{2} & 0 \\
 0 & -2 \mathrm{i} Z_3 \sin \frac{k_3}{2} & 0 & 0 & 0 & -2 \mathrm{i} \bar{Z}_2 \cos \frac{k_2}{2} & w_1 & 0 & 0 & 0 & 2 \bar{\tau} \cos \frac{k_3}{2} & -2 \tau  \sin \frac{k_2}{2} \\
 -2 \mathrm{i} \bar{Z}_3 \sin \frac{k_3}{2} & 0 & 0 & 0 & \mathrm{i} w_2 & 0 & 0 & 0 & 2 Z_1 \cos \frac{k_1}{2} & 2 \tau  \cos \frac{k_3}{2} & 0 & -2 \bar{\tau} \sin \frac{k_1}{2} \\
 0 & 0 & -\mathrm{i} w_3 & -2 \mathrm{i} Z_2 \cos \frac{k_2}{2} & 0 & 0 & 0 & 2 \bar{Z}_1 \cos \frac{k_1}{2} & 0 & -2 \bar{\tau} \sin \frac{k_2}{2} & -2 \tau  \sin \frac{k_1}{2} & 0 
\end{pmatrix}}.
\label{big_bloch_hamiltonian}
\end{equation}
\subsubsection{Electronic dispersion in the AFM phase}

Consider the low-symmetry AFM state $\textbf{W}=(W,0,0)$, $\textbf{w}=(w,0,0)$, $\boldsymbol{\Phi}=(0,\Phi,\Phi')$, which corresponds to phase 5 in Table \ref{tab:GS_table}. Note that the high-symmetry AFM state corresponding to phase 2 can be obtained by setting $\Phi'=\Phi$. Plugging in the values of $Z_i$ yields $Z_1$ real and $Z_{2,3}$ imaginary. The gauge choice of Eq. (\ref{big_bloch_hamiltonian}) ensures that the following prefactors are all real: 
\begin{equation}
    e^{\mathrm{i}\alpha_1}Z_1\text{, }e^{\mathrm{i}\alpha_2}Z_2\text{, }e^{\mathrm{i}\alpha_3}Z_3\text{, }e^{\mathrm{i}\alpha_1}w_1\in\mathbb{R}
\end{equation}
\begin{figure*}
\includegraphics[width=1.0\columnwidth]{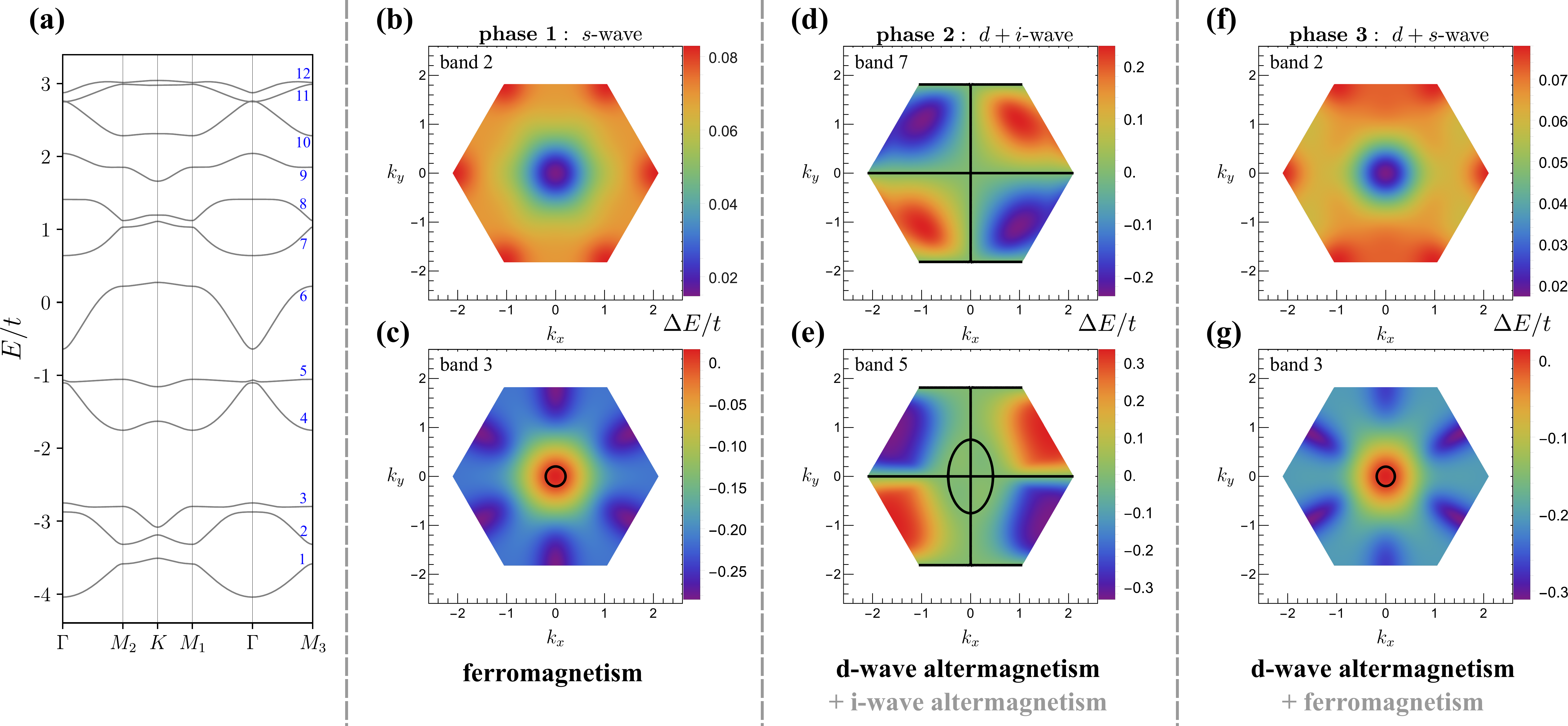}\caption{\textbf{(a)} Tight-binding band structure of the parent CDW state with $\textbf{W}=(0.5,0.5,0.5)$ and the corresponding band labels. \textbf{(b)}-\textbf{(g)} Color plots of the $\textbf{k}$-dependent spin-splitting $\Delta E=E_\uparrow - E_\downarrow$ on the mini-BZ (BZ') for selected bands and magnetic states. The bands are labeled $1$-$12$ based on their ordering with increasing energy in the parent state (in blue). The results are obtained from diagonalizing $\tilde{H}(\textbf{k})$ with $\lambda_\text{soc}=0.2t$ for different CDW-LC mixed states corresponding to phases $1$, $3$, and $4$ in Table \ref{tab:GS_table}. The black curves correspond to nodes ($\Delta E=0$). The top row only exhibits symmetry-enforced nodes (if they exist) whereas the bottom row also exhibits accidental nodes. \textbf{(b, c)}: Phase $1$ configuration with $\textbf{W}=\textbf{w}=\boldsymbol\Phi=(0.5,0.5,0.5)$, corresponding to $D_{6h}$-preserving FM (i.e., $s$-wave). \textbf{(d, e)}: Phase $3$ configuration with $\textbf{W}=\textbf{w}=(0.5,0.4,0.5)$ and $\boldsymbol\Phi=(0.5,0,-0.5)$, corresponding to $d$-wave AM mixed with $i$-wave AM. \textbf{(f, g)}: Phase $4$ configuration with $\textbf{W}=\textbf{w}=\boldsymbol\Phi=(0.5,0.4,0.5)$, corresponding to $d$-wave AM mixed with FM.}\label{fig:BZ_splitting}
\end{figure*}

It follows that the only imaginary content of the Bloch Hamiltonian is proportional to $\Imaginary \tau=-\lambda_\text{soc}$. Because of its Hermiticity, $\tilde{H}(\textbf{k})$ and its complex conjugate have the same eigenvalues for all $\textbf{k}$. Thus, since complex conjugating the Bloch Hamiltonian $\tilde{H}(\textbf{k})$ is equivalent to flipping the spin, this result implies that the bands are spin-degenerate throughout the reduced Brillouin zone in the AFM state to all orders in $\lambda_\text{soc}$. We plot the spin-degenerate bands for phase 2 numerically in Fig. 3(b) of the main text, using the parameters $\lambda_\text{soc}=0.4t$, $\textbf{W}=\textbf{w}=(0,0.5,0)$, and $\boldsymbol{\Phi}=(0.4,0,0.4)$.

\subsubsection{Electronic dispersion in the FM and AM states}

To explicitly obtain the spin-splitting of the energy dispersions  $\Delta E(\textbf{k})=E_\uparrow(\textbf{k})-E_\downarrow(\textbf{k})$  that we derived from group theory in Section ID, we compute the tight-binding spectrum of Eq. (\ref{momentum_space_hamiltonian_2}) numerically for the phase $1$ (FM), phase $3$ ($d$-wave AM), and phase $4$ ($d$-wave AM + FM) of Table \ref{tab:GS_table}. The results for phase $1$ (FM) and phase $3$ ($d$-wave AM) correspond to main text figures 3(a) and 3(c-d), respectively. The parameters used were $\lambda_\text{soc} =0.4t$, and $\textbf{W}=\textbf{w}=-(0.5,0.5,0.5)$, $\boldsymbol{\Phi}=(0.4,0.4,0.4)$ for the FM case, and $\textbf{W}=\textbf{w}=-(0.5,0.4,0.5)$, $\boldsymbol{\Phi}=(0.4,0,-0.4)$ for the AM case.

Our procedure to obtain these results is as follows. As discussed above, $\tilde{\mathcal{H}}$ is block-diagonalized in the spin basis and taking $\tau\to\bar\tau$ ($\lambda_\text{soc}\to-\lambda_\text{soc}$) is equivalent to flipping the spin. We therefore fix $\lambda_\text{soc}$ and compute the corresponding $\textbf{k}$-dependent eigenvalues of the $12$-by-$12$ matrix $\tilde{H}(\textbf{k})$ and compare with those of $-\lambda_\text{soc}$. This yields, respectively, the spin-up and spin-down dispersions $E^n_\uparrow(\textbf{k})$ and $E^n_\downarrow(\textbf{k})$, where $n$ is the band index. A color-plot of the momentum-dependent spin-splitting $\Delta E^n(\textbf{k})\equiv E^n_\uparrow(\textbf{k})-E^n_\downarrow(\textbf{k})$ for fixed $n$, $Z_i$, $w_i$, and $\lambda_\text{soc}$ is plotted on the mini Brillouin zone in Fig. \ref{fig:BZ_splitting}. In this figure, while the first column shows the band dispersion in phase 0 (i.e., CDW without LC order), each of the other columns corresponds to a $Z_i$, $w_i$ configuration (given in the caption) that realizes one of the three states phase $1$, phase $3$, and phase $4$. Within each column, each row refers to a different band with all other parameters held the same. Clearly, the form of $\Delta E^n(\textbf{k})$ in each column agrees with the results obtained in Section ID based solely on symmetry. Interesting, the results show that, within the same microscopic model, one can use the band index to tune between spin-split bands that exhibit no accidental nodes (top row) and those that host accidental nodes (bottom row).  This means that the phenomenological constants introduced in Eq. (\ref{GS1_splitting})-(\ref{GS4_splitting}) are explicit functions of the tight-binding parameters, $c_i=c_i(\textbf{Z},\textbf{w},\lambda_\text{soc},n)$.

Notably, the presence of an enlarged symmetry in the limit of no site order ($\textbf{w}=0$) makes $\tilde{H}(\textbf{k})$ amenable to analytical results at the $\Gamma$-point. Recall that spin-splitting in the $\Gamma$-point is only allowed in the presence of FM order. Without CDW site order, $\tilde{H}(\textbf{k}=0)$ assumes a block-diagonal form with a $3$-by-$3$ $H_a$ block and a $9$-by-$9$ block $H_b$:
\begin{equation}
    \tilde{H}(\textbf{k}=0)|_{\textbf{w}=0}=\begin{pmatrix}H_a & \textbf{0}^T\\ \textbf{0} & H_b\end{pmatrix}
\end{equation} where 
\begin{equation}
    H_a = -2\begin{pmatrix}
        0 & \bar\tau & \tau \\\tau & 0 & \bar\tau \\ \bar\tau & \tau & 0
    \end{pmatrix}\quad\text{and}\quad H_b=2 \begin{pmatrix}
        0 & 0 & 0 & 0 & \mathrm{i}Z_3 & 0 & 0 & 0 & \mathrm{i} \bar{Z_2} \\
        0 & 0 & -\bar\tau & \mathrm{i} \bar{Z_3} & 0 & 0 & 0 & 0 & 0 \\
        0 & -\tau & 0 & 0 & 0 & 0 & \mathrm{i}Z_2 & 0 & 0 \\
        0 & -\mathrm{i}Z_3 & 0 & 0 & 0 &\tau & 0 & 0 & 0 \\
        -\mathrm{i}\bar{Z_3} & 0 & 0 & 0 & 0 & 0 & 0 & 0 & Z_1 \\
        0 & 0 & 0 & \bar\tau & 0 & 0 & 0 &\bar{Z_1} & 0 \\
        0 & 0 & -\mathrm{i}\bar{Z_2} & 0 & 0 & 0 & 0 &\bar\tau & 0 \\ 
        0 & 0 & 0 & 0 & 0 &Z_1 & \tau & 0 & 0 \\
        -\mathrm{i}Z_2 & 0 & 0 & 0 &\bar{Z_1} & 0 & 0 & 0 & 0
    \end{pmatrix}.
\end{equation}
Surprisingly, only $H_b$ depends on $\textbf{W}$ and $\boldsymbol{\Phi}$, and furthermore, only six of the twelve $\Gamma$-point levels split in the presence of FM order. The splitting of the remaining six bands are restored when $\textbf{w}\neq 0$. These facts imply that the assumption of no site order makes the Hamiltonian have more symmetries that expected. Nevertheless, given the fact that the Hamiltonian is analytically tractable in this limit, we proceed to determine the relationship between the $\mathbf{k}=0$ splitting  $\Delta E^n$ of those six bands in the FM phase and the FM composites derived in Section IB. 

While analytical expressions for the $\Gamma$-point energy levels cannot be obtained, we gain useful insights by analyzing the characteristic polynomial $p(E)$ of the nontrivial block $H_b$: 
\begin{equation}
    p(E)=\det(E\mathbb{I}-H_b)=p_0+p_1E+p_2E^2+p_3E^3+p_4E^4+p_5E^5+p_6E^6+p_7E^7+p_8E^8+E^9
\end{equation}
When written in terms of the complex variables $Z_i=W_i+\mathrm{i}\Phi_i$, the coefficients take a rather simple form:
\begin{equation}
    \begin{split}
    p_0 &= -1024|Z_1Z_2Z_3-\tau^3|^2\Real(Z_1Z_2Z_3)\\
    p_1 &= -256|Z_1Z_2Z_3-\tau^3|^2\big(|Z_1|^2+|Z_2|^2+|Z_3|^2\big)\\
    p_2 &= 256 \Real(Z_1Z_2Z_3)\bigg[\sum_i\big(|Z_i|^2+|\tau|^2\big)\big(|Z_j|^2+|\tau|^2\big)\bigg]\\
    p_3 &= 64\Bigg\{|Z_1Z_2Z_3-\tau^3|^2-3|Z_1Z_2Z_3|^2\\&+ \sum_i \big[|Z_i|^4(|Z_j|^2+|Z_l|^2+|\tau|^2)+|\tau|^2(2|Z_j|^2|Z_l|^2+3|\tau|^2|Z_i|^2) \big]\Bigg\}\\
    p_4 &= -64 \Real(Z_1Z_2Z_3)\sum_i\big(|Z_i|^2+|\tau|^2\big)\\
    p_5 &= -16\sum_i\big(|Z_i|^4+4|\tau|^2|Z_i|^2+3|Z_j|^2|Z_l|^2+|\tau|^4\big)\\
    p_6 &= 16 \Real(Z_1Z_2Z_3)\\
    p_7 &= 4\sum_i \big(2|Z_i|^2+|\tau|^2 \big)\\
    p_8 &= 0.
    \end{split}
    \end{equation}
We notice that the only coefficients that change under $\tau\to\bar\tau$ are $p_0$, $p_1$, and $p_3$, as they depend on more than just the magnitude of $\tau$, i.e., through $|Z_1Z_2Z_3-\tau^3|$. This term, when expanded to linear order in SOC, gives 
\begin{equation}
    |Z_1Z_2Z_3-\tau^3|-|Z_1Z_2Z_3-t^3|\propto \lambda_\text{soc}\Imaginary(Z_1Z_2Z_3)+\mathcal{O}(\lambda_\text{soc}^2).
\end{equation}
We stress that the roots of polynomials are regular functions of its coefficients whenever the roots are strictly non-degenerate. Thus, we apply Taylor's theorem and expand $E_{\lambda_\text{soc}}^n$ and $E_{-\lambda_\text{soc}}^n$ in $\lambda_\text{soc}$ to obtain the $\Gamma$-point spin-splitting to lowest order in SOC:
\begin{equation}
    \Delta E^n\propto\lambda_\text{soc}\Imaginary(Z_1Z_2Z_3)=\lambda_\text{soc}(\Phi_1 W_2 W_3+\Phi_2W_3 W_1+\Phi_3W_1W_2-\Phi_1\Phi_2\Phi_3) = \lambda_\text{soc}(\mathcal{M}_2-\mathcal{M}_1)
\end{equation}
where the constant of proportionality depends on $n$ and $Z_i$. We have also recognized the trilinear terms $\Phi_1 W_2 W_3+\Phi_2W_3 W_1+\Phi_3W_1W_2$ and $\Phi_1 \Phi_2 \Phi_3$ as the higher-order FM multipoles $\mathcal{M}_2$ and $\mathcal{M}_1$ defined in Eq. (\ref{eq:M12}), respectively. Since they have the same symmetry as $\mathcal{M}$, they must be proportional to it, yielding the expected result that the $\Gamma$-point splitting depends on the magnetization and on the SOC:  
\begin{equation}
    \Delta E^n\propto\lambda_\text{soc}\mathcal{M}.
\end{equation}
The same results hold in the presence of site-order, although this analysis becomes more complicated.
\bibliography{AMpaper_references}